\documentclass[a4paper,11pt]{article}
\pdfoutput=1
\usepackage{a4wide}
\usepackage[latin1]{inputenc}
\usepackage{dsfont}
\usepackage{amsfonts}
\usepackage{amsmath}
\usepackage{amssymb}
\usepackage{amsthm}
\usepackage{graphicx}
\usepackage{booktabs}
\usepackage[numbers, sort&compress]{natbib}
\allowdisplaybreaks[1]
\usepackage{color}
\usepackage{ulem}

\usepackage[small,bf]{caption}
\def\sgn{\mathop{\rm sgn}}
\newcommand{\ci}{\text{i}}

\begin{document}

\begin{titlepage}

\vspace*{-15mm}
\begin{flushright}
MPP-2011-43\\
SISSA 17/2011/EP
\end{flushright}
\vspace*{0.7cm}

\begin{center}
{\LARGE {\bf From Flavour to SUSY Flavour Models}}\\[8mm]

Stefan Antusch$^{\star\dagger}$\footnote{Email: \texttt{antusch@mppmu.mpg.de}},~
Lorenzo Calibbi$^{\dagger}$\footnote{Email: \texttt{calibbi@mppmu.mpg.de}},~
Vinzenz Maurer$^{\dagger}$\footnote{Email: \texttt{maurer@mppmu.mpg.de}}~
and
Martin Spinrath$^{\S}$\footnote{Email: \texttt{spinrath@sissa.it}}
\end{center}

\vspace*{0.50cm}

\centerline{$^{\star}$ \it
Department of Physics, University of Basel,}
\centerline{\it
Klingelbergstr.\ 82, CH-4056 Basel, Switzerland}

\vspace*{0.4cm}

\centerline{$^{\dagger}$ \it
Max-Planck-Institut f\"ur Physik (Werner-Heisenberg-Institut),}
\centerline{\it
F\"ohringer Ring 6, D-80805 M\"unchen, Germany}

\vspace*{0.4cm}

\centerline{$^{\S}$ \it
SISSA/ISAS and INFN,}
\centerline{\it
Via Bonomea 265, I-34136 Trieste, Italy }
\vspace*{1.20cm}

\begin{abstract}
\noindent
If supersymmetry (SUSY) will be discovered, successful models of flavour not only have to provide an explanation of the flavour structure of the Standard Model fermions, but also of the flavour structure of their scalar superpartners. We discuss aspects of such ``SUSY flavour'' models, towards predicting both flavour structures, in the context of supergravity (SUGRA). We point out the importance of carefully taking into account SUSY-specific effects, such as 1-loop SUSY threshold corrections and canonical normalization, when fitting the model to the data for fermion masses and mixings. This entangles the flavour model with the SUSY parameters and leads to interesting predictions for the sparticle spectrum. We demonstrate these effects by analyzing an example class of flavour models in the framework of an $SU(5)$ Grand Unified Theory with a family symmetry with real triplet representations. For flavour violation through the SUSY soft breaking terms, the class of models realizes a scheme we refer to as ``Trilinear Dominance'', where flavour violation effects are dominantly induced by the trilinear terms.

\end{abstract}

\end{titlepage}

\tableofcontents

\newpage

\section{Introduction}

The flavour puzzle is one of the biggest open questions in particle physics. Currently, one may subdivide it into three main parts: the puzzle associated with the masses and mixing properties of the quarks and charged leptons, the new puzzles that have been added with the discovery of neutrino masses and mixing, as well as the issues connected to CP violation.

Regarding the masses of quarks and charged leptons it is tantalizing that down-type quarks and charged leptons show a similar hierarchical pattern, which differs substantially from the pattern of the up-type quark masses. In the lepton sector, in addition to the smallness of the neutrino masses the observation of large flavour mixing, in contrast to the small quark mixing, has to be explained. Concerning CP violation, there is on the one hand the strong CP problem, but on the other hand the observation that the unitarity triangle is, at least, almost right-angled, i.e.\ $\alpha \approx 90^\circ$. If supersymmetry (SUSY) or any other kind of new physics will be discovered at the LHC, this would add another ``dimension'' to the flavour puzzle. Any kind of new physics has to face severe constraints from flavour physics, e.g.\ flavour changing neutral currents in the quark sector (FCNCs), lepton flavour violation (LFV) as well as electric dipole moments (EDMs). In the case of SUSY this is often referred to as the ``SUSY flavour puzzle''.

Recently a lot of effort has been made to understand the flavour structure of quarks and leptons using non-Abelian family symmetries such as $A_4$, $S_4$, $SO(3)$, $SU(3)$, and so forth (see, e.g.\  \cite{Altarelli:2010gt}), where the three families of SM fermions can be unified in representations of the family symmetry. For reviews with extensive lists of references, we refer the reader to \cite{FlavourRev}.
One strength of such non-Abelian family symmetries is that they can provide an explanation of the observed close to tri-bimaximal mixing in the lepton sector. Most of these  models are also formulated in a SUSY context, although this is often not worked out in any detail. In many works only the superpotential relevant for the vacuum alignment of the family symmetry breaking Higgs fields, the ``flavons'', and the fermion masses and mixing angles is presented, without considering the consequences for the SUSY breaking sector. It is one main aim of this paper to highlight the entanglement between the SM and the SUSY sector when a careful model analysis is performed. 

Family symmetries (in particular non-Abelian families discussed towards explaining tri-bimaximal lepton mixing) can indeed constrain the structure of the sfermion mass matrices such that compatibility with FCNC constraints or LFV is improved, see, e.g. \cite{Abel:2001cv, Ross-Vives, Ross:2004qn,Antusch:2007re,Olive:2008vv, su3edms,Feruglio:2009hu,Calibbi:2010rf,Lalak:2010bk}. Several aspects regarding the connection between family symmetries and the SUSY flavour structure have been discussed in the literature: For example the role of induced flavon F-terms has been analysed in \cite{AKMR,Feruglio:2009iu}. Canonical normalisation effects in family symmetry models in supergravity, and the dependence of their relevance on the messenger sector of the models, have been studied in \cite{Antusch:2007vw}. In another recent study, the impact of the messenger sector on the predictivity of family symmetries for the soft SUSY breaking parameters has been highlighted \cite{Kadota:2010cz}. 
Furthermore, in combination with family symmetries, supersymmetric Grand Unified Theories (GUTs) are an attractive framework for trying to solve the flavour and SUSY flavour puzzles, since they provide a link between the different (s)fermion species, i.e.\ between quarks and leptons respectively squarks and sleptons. 

Many studies in the literature focus either on explaining the flavour structure of the fermion masses and mixing parameters \emph{or} on the SUSY flavour puzzle, but not on both at the same time. However, if SUSY will be discovered, it is clear that  successful models of flavour will have to provide an explanation for both,  the flavour structure of the Standard Model fermions \emph{and} the flavour structure of their scalar superpartners. 
In this paper, we discuss aspects of such ``SUSY flavour'' models, which aim at predicting both flavour structures simultaneously, in the context of supergravity (SUGRA).
We point out the importance of carefully taking into account SUSY-specific effects, such as 1-loop SUSY threshold corrections and canonical normalization effects,  when fitting the model to the data for fermion masses and mixings. These effects entangle the flavour model with the SUSY parameters and lead to interesting predictions for the sparticle spectrum as well as, for instance, for the neutrino parameters. In addition, family symmetries introduced to explain the flavour structure of the Standard Model fermions can also make predictions testable in future flavour experiments. We demonstrate these effects by analyzing an example class of flavour models in the framework of $SU(5)$ Grand Unified Theories with a family symmetry with real triplet representations. Thereby we benefit from the GUT Yukawa relations $y_\mu/y_s = 6$ and $y_\tau/y_b = 3/2$ two of us proposed in \cite{Antusch:2009gu}. From there we derive, in a semi-analytical as well as in a numerical analysis, the constraints on the SUSY parameters and the predictions for FCNCs, LFV and EDMs. In the class of SUSY flavour models considered as example, flavour violation through the SUSY soft breaking terms is dominated by the trilinear terms. We discuss the possible signatures of such a scheme of ``Trilinear Dominance'' \footnote{We note that the scenario we consider here differs from the one proposed in \cite{Crivellin:2008mq}, 
where it is assumed that the fermion mixing is generated entirely by radiative effects
especially from the trilinear terms. In our case, the SM and SUSY flavour structures are both generated from 
family symmetry breaking. Trilinear dominance means that the misalignment between the Yukawa 
matrices and the A-term matrices dominates the new physics flavour violation effects.}.

The paper is organized as follows: In Sec.\ \ref{sec:Model} we first define the class of models we will consider as an example. At this point the model is not yet supersymmetric, i.e.\  we specify only the fields necessary to describe the SM fermion masses and mixing angles. In Sec.\ \ref{sec:SUSY} we discuss the relevant SUSY specific effects, like canonical normalization and SUSY threshold corrections. There, we also discuss the structure of the soft SUSY breaking terms, first in the flavour basis and then also in the SCKM basis. The fourth section is devoted to a semi-analytical analysis, where we describe an approximate procedure to extract the low energy flavour observables from the high scale model. We also show how, from a fit of the low energy fermion masses and mixing angles to the data, constraints on the SUSY spectrum arise. In Sec.\ \ref{sec:num}, we discuss the results of a MCMC analysis, which supports our semi-analytical results. We also present there phenomenological consequences for the SUSY spectrum and low-energy
observables. The last section contains a summary and our conclusions.

\section{An Example Class of Flavour Models} \label{sec:Model}

Before going to the SUSY specific ingredients we will first describe in general terms the class of flavour models we are going to consider as an example for all the SUSY specific effects. We emphasize that our analysis is not restricted to this special class of models. In contrary it should be applied to any given SUSY model of flavour to check its validity. As we will see it is necessary to include SUSY effects to check if a correct description of SM fermion masses and mixing angles can be obtained.

\subsection{Basic Structure of the Example Class of Models}
Our starting point for describing the class of flavour models, to be discussed as an example in the next sections, is a Grand Unified Theory (GUT) with gauge group $SU(5)$ plus a non-Abelian family symmetry group $G_F$ with real triplet representations such as, for example, $A_4$ or $SO(3)$. For the matter fields of the SM fields $SU(5)$-unification implies that they are unified in the $\bar{\mathbf{ 5}}$ representations $F_i$ and in the $\mathbf{10}$ representations $T_i$, where $i=1,2,3$ labels the three families, as follows:
\begin{equation}
F_i  =  \begin{pmatrix}
      d_R^{c} & d_B^{c} & d_G^{c} & e &-\nu
       \end{pmatrix}_i \end{equation}
and
\begin{equation}
T_i   = \frac{1}{\sqrt{2}}
        \begin{pmatrix}
        0 & -u_G^{c} & u_B^{c} & -u_{R} & -d_{R} \\
        u_G^{c} & 0 & -u_R^{c} & -u_{B} & -d_{B} \\
        -u_B^{c} & u_R^{c} & 0 & -u_{G} & -d_{G} \\
        u_{R} & u_{B} & u_{G} & 0 & -e^c \\
        d_{R} & d_{B} & d_{G} & e^c & 0
        \end{pmatrix}_i .
\end{equation}
We will assume that the three fields $F_i$ are components of a triplet $F$ under the family symmetry $G_F$.  The other three fields $T_i$ remain singlets under $G_F$. As we will see, this allows the top mass to emerge at renormalizable level, while the masses of the other lighter fermions will be generated from effective operators after family symmetry breaking. In addition to the SM fermions, we add two right-handed neutrinos $N_1,N_2$ to the spectrum, which are singlets under the family symmetry, in order to allow for two massive light neutrinos via the seesaw mechanism as observed in neutrino oscillations. Note that this is a minimal setup and implies that one of the light neutrinos is massless by construction. It is straightforward to extend the model to include a third right-handed neutrino, to give mass to all three light neutrinos.  

As mentioned above, apart from the top quark mass all masses and thus also the mixing parameters of the fermions arise after the family symmetry (and the EW symmetry) gets broken. Following \cite{Antusch:2010es} we will assume that the family symmetry is broken by flavour-Higgs fields (the flavons) $\phi_{i}$ in triplet representations of $G_F$, acquiring vevs in the following directions in flavour space\footnote{We note that in explicit models one might probably split up $\langle \tilde\phi_{2} \rangle$ into two flavons (cf.\ Appendix B of \cite{Antusch:2010es}), one with a purely real and another one with a purely imaginary vev. A mechanism for realising such vacuum alignments is presented in \cite{Antusch:2011sx}.}:
\begin{gather} 
\langle \phi_{1} \rangle \propto \begin{pmatrix}  0 \\  1\\  - 1  \end{pmatrix}  \;, \quad
\langle \phi_{2} \rangle \propto \begin{pmatrix} 1 \\  1 \\ 1  \end{pmatrix} \;, \quad
\langle \phi_{3} \rangle   \propto \begin{pmatrix}  0 \\  0\\  1  \end{pmatrix} \;, \quad
\langle \tilde\phi_{2} \rangle  \propto \begin{pmatrix}  0 \\  - \mathrm{i} \\  w  \end{pmatrix} \;.
\end{gather}
The least understood and most difficult-to-test part of current flavour models with non-Abelian family symmetries is their so-called ``vacuum alignment'' sector. This is the part of the theory which addresses the question why the flavons point in these specific directions in flavour space. In this paper we will not enter this part of the flavour models in much detail. However, we will come back later to one (at least in principle) testable feature of the vacuum alignment sector in supersymmetric models, namely the size of the induced F-terms for the flavons. Finally, we remark that the scale of family symmetry breaking can in principle vary over a wide range. In the following, we will assume for definiteness that the family symmetry is broken close to the GUT scale. 

We will now describe the effective Lagrangian of the model class. For its origin from a renormalizable theory we refer the interested reader to Sec.\ \ref{messengers} where we discuss the messenger field sector and sketch the additional discrete symmetries of the setup. In Sec.\ \ref{Yukawa} we will describe the resulting Yukawa matrices and the mass matrix of the right-handed neutrinos. A reader mainly interested in the phenomenological analysis may skip Sec.\ \ref{messengers}.  

The effective Lagrangian of the class of models considered here has the following structure:
\begin{align}
-\mathcal{L}^{\text{eff}}_{Y_{d,l}} &= \!\sqrt{2}\,F H_{24} \!\left(\frac{\phi_{1}}{M_{1}M_{1}^\prime} T_1 + \frac{\phi_{2}}{M_{2}M_{2}^\prime} T_2 +  \frac{\phi_{3}}{M_3 M_3^\prime} T_3 \right) \bar{H}_5 +\! \sqrt{2}\,
T_2 \frac{H_{24}}{\tilde{M}_2} \frac{\tilde{\phi}_{2}}{\tilde{M}_{2}^\prime} F \bar{H}_{5} + {\rm H.c.}\;, \label{Eq:Yl} \\
-\mathcal{L}^{\text{eff}}_{Y_u} &= \frac{1}{4}\left( y_3 T_3^2 + \frac{\phi_{2}^2}{M^u_{2}M^{\prime u}_{2}} T_2^2 + \frac{\phi_{1}^2}{M^u_{1}M^{\prime u}_{1}} T_1^2  \right) H_5 + {\rm H.c.}\;, \label{Eq:Yu} \\
-\mathcal{L}^{\text{eff}}_{Y_\nu} &= F \left(\frac{\phi_{1}}{M_1} N_1 +  \frac{\phi_{2}}{M_2} N_2 \right) H_5 + {\rm H.c.}\;, \label{Eq:Ynu} \\
-\mathcal{L}^{\text{eff}}_{M_{\nu^c}} &= \frac{\phi_{1}^2}{M^{\prime u}_{1}} N_1^2 + \frac{\phi_{2}^2}{M^{\prime u}_{2}} N_2^2 + {\rm H.c.}\;. \label{Eq:MR}
\end{align}
$H_{24}$ stands for a Higgs field in the $\mathbf{24}$ representation that breaks $SU(5)$ to the SM, and $\bar{H}_5$ and $H_5$ stand for Higgs fields that contain the MSSM Higgses $H_d$ and $H_u$ in their SM-doublet components. The Higgs fields are singlets under the family symmetry.
To shorten notation, $\mathcal{O}(1)$ coefficients have been absorbed (without loss of generality) in the mass scales in the denominators, which correspond to the masses of heavy ``messenger fields'' as will be discussed below. As one can see, only ratios of flavon vevs over the messenger masses enter the Lagrangian, reflecting the fact that the family symmetry breaking scale is a priori undetermined.

\subsection{Messenger Fields and Discrete Symmetries}\label{messengers}
Some properties of the effective Lagrangian are determined by the family symmetry $G_F$. For example, since $F$ is a triplet under $G_F$ it has to appear together with a flavon in the terms in $\mathcal{L}^{\text{eff}}_{Y_{d,l}} $ which lead to the down-type and charged lepton Yukawa matrices. This implies that the rows of $Y_d$ and the columns of $Y_e$ will be generated dynamically by the flavon vevs. Other properties of the effective Lagrangian, for example that the top quark Yukawa coupling appears at the renormalizable level while all other up-type Yukawa couplings do not, requires additional symmetries and the knowledge of the ``messenger sector'', i.e.\ of the fields which are interchanged to generate the effective interactions. In summary, the class of models we like to outline here has the following features:
\begin{itemize}
\item The rows of $Y_d$ and the columns of $Y_e$ will be generated dynamically by the flavon vevs with the rule that
$T_1$ couples only to $\phi_{1}$, $T_2$ couples only to $\phi_{2}$ and $\tilde{\phi}_{2}$, and $T_3$ couples only to $\phi_3$.
\item The Yukawa coupling ratios $y_\tau/y_b$ and $y_\mu/y_s$ at the GUT scale will be predicted as $3/2$ and $6$, respectively, by Clebsch--Gordan coefficients from GUT symmetry breaking. As shown in \cite{Antusch:2009gu} this requires a moderate or large value of
$\tan \beta$.
\item At leading order $Y_u$ will be diagonal.
\item Left-handed neutrino masses will be generated via a see-saw type I mechanism with two right-handed neutrinos.
\item The mixing in the neutrino sector is of almost tri-bimaximal form.
\item We consider spontaneous CP violation from the flavon vevs (see, e.g.\ \cite{Ross:2004qn,Antusch:2007re}), in our specific case from $\langle \tilde\phi_{2} \rangle$ only.
\end{itemize}

\begin{figure}
\centering
\includegraphics[scale=0.69]{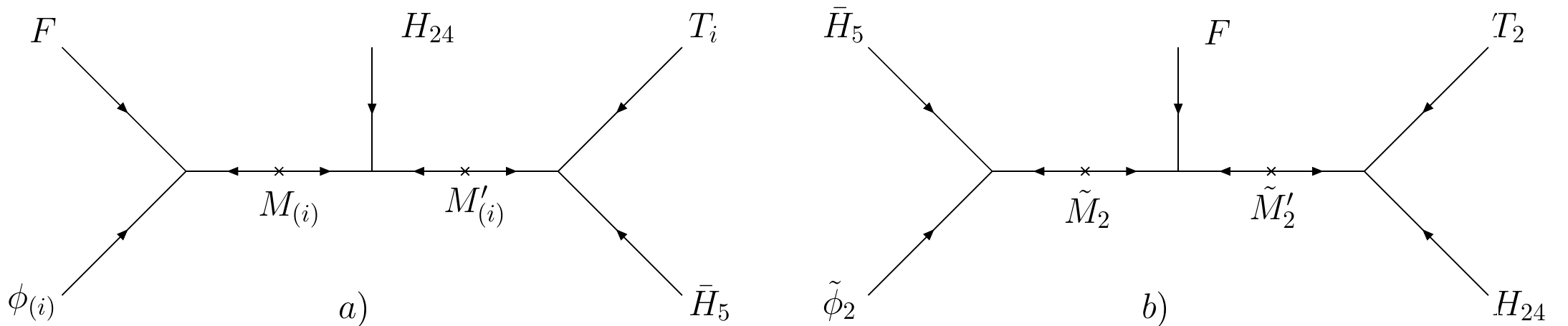}
\caption{Supergraph diagrams inducing the effective superpotential operators for the down-type quarks and charged leptons. \label{Fig:messenger_d}}
\end{figure}

The first point listed above is realized, for example, when the $T_i$ as well as the associated flavons carry some charge under a specific additional discrete symmetry, for instance a $\mathbb{Z}_n$ symmetry. In addition, we have to make sure that 
the ratios among Yukawa couplings mentioned above are reproduced by correct contractions of the $SU(5)$ indices of the Higgs 
and matter representations. 
This can be achieved if the effective operators are generated by integrating out the appropriate messenger fields \cite{Antusch:2009gu}. 
The relevant operators for the down-quark and charged lepton sector are, for example:
\begin{align}
\frac{1}{M^2} (F \phi_{3} )_{\bar{5}} H_{24} ( \bar H_5 T_3 )_{5}
\end{align}
leading to $y_\tau/y_b = 3/2$ and
\begin{align}
\frac{1}{M^2} (\bar H_5 \tilde\phi_{2} )_{\bar{5}} F  ( H_{24} T_2 )_{10}
\end{align}
leading to $y_\mu/y_s = 6$ at the GUT scale. Here the subscript refers to the used $SU(5)$ index contraction and $M$ represents generically the mass scale of the messenger fields, which will be specified below. $H_{24}$ and $\bar H_5$ are Higgs fields in the $\mathbf{24}$ and in the $\bar{\mathbf{ 5}}$ representations of $SU(5)$ respectively. 
$H_{24}$ can be the usual Higgs responsible for the breaking of $SU(5)$ to the SM gauge group by acquiring a vev $\langle H_{24} \rangle = v_{24} \: \text{diag} (1,1,1,-3/2,-3/2)$.
Clearly, the messenger fields, which generate these operators, can be distinguished by their charges under the additional discrete symmetries which distinguish between $T_2$ and $T_3$.
The diagrams generating the Yukawa couplings for down-type quarks and charged leptons are displayed in Fig.~\ref{Fig:messenger_d}, where the masses
of the messenger fields are also shown.

\begin{figure}
\centering
\includegraphics[scale=0.69]{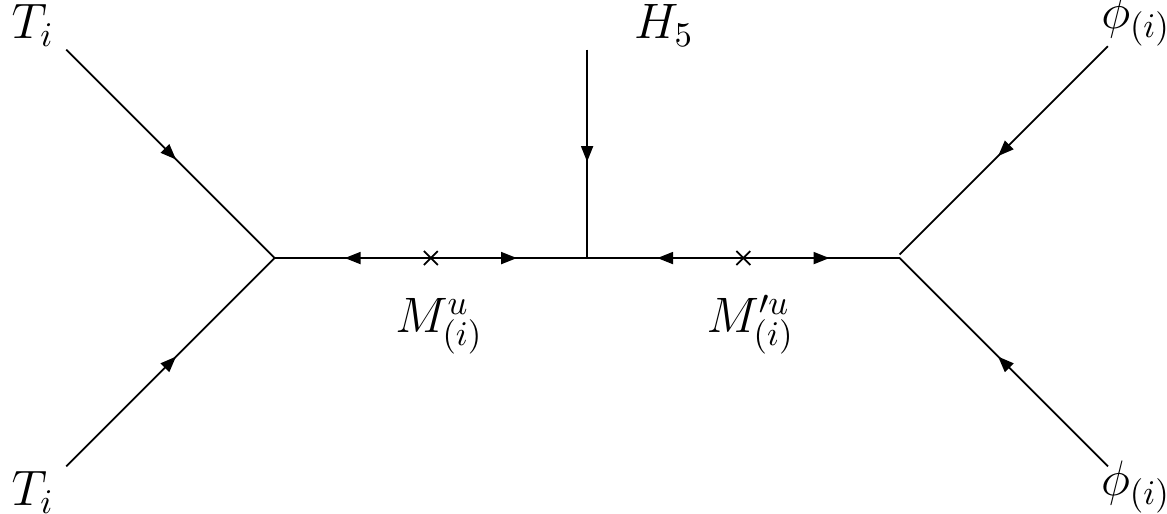}
\caption{Supergraph diagrams inducing the effective superpotential operators for the up-type quarks. \label{Fig:messenger_u}}
\end{figure}

\begin{figure}
\centering
\includegraphics[scale=0.69]{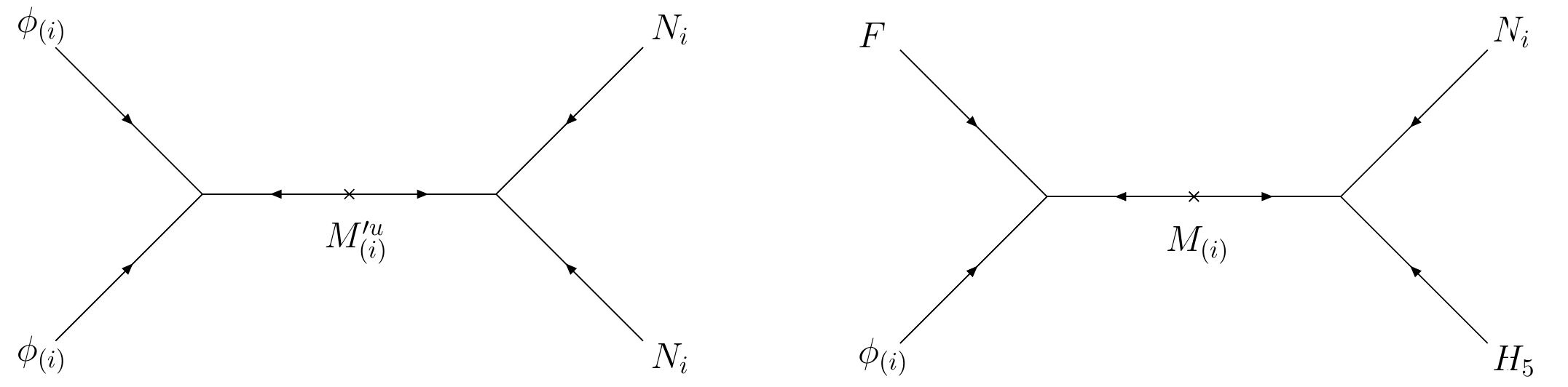}
\caption{Supergraph diagrams inducing the effective superpotential operators for the neutrino sector. \label{Fig:messenger_n}}
\end{figure}

If we would introduce at least one discrete symmetry for each generation and the messenger fields generically carry charge under exactly one of these symmetries, it follows that the up-quark Yukawa matrix is generated with a diagonal structure.
Specifically, we will arrange that only the top quark mass is generated at the renormalizable level, by associating $T_3$ to a $\mathbb{Z}_2$ symmetry, for instance, while the charm quark and up quark masses are generated only after family symmetry breaking, by associating $T_2$ and $T_1$ to larger symmetries as, e.g.\ $\mathbb{Z}_n$ symmetries with $n > 2$. The relevant operator for the up-quark mass can then be
\begin{align}
\frac{1}{M^2} \left(T_1 T_1\right)_{\bar{5}} H_{5} \left(\phi_{1}\phi_{1}\right)_{1} \;,
\end{align}
generated via diagrams as shown in Fig.~\ref{Fig:messenger_u}. Similar considerations can be made for the neutrino sector, where now the RH neutrinos $N_i$ carry the same charges as the $T_i$ to associate them with the flavons. Majorana masses for the RH neutrinos and neutrino Yukawa couplings are then generated for instance by the diagrams shown in Fig.~\ref{Fig:messenger_n}.

In summary, we will consider models with auxiliary discrete symmetries to separate each of the $T_i$, leading to the following pairings between matter and flavon fields:
\begin{equation}
\begin{split}\label{Eq:Pairing}
T_1, N_1 &\longleftrightarrow \phi_{1} \;, \\
T_2, N_2 &\longleftrightarrow \phi_{2} \;, \\
T_2 &\longleftrightarrow \tilde{\phi}_{2} \;, \\
T_3 &\longleftrightarrow \phi_{3} \;.
\end{split}
\end{equation}
We note that $F$ is not paired to a particular flavon since it is a triplet under the family symmetry and contains all three generations of the five-dimensional matter representations.
Basically the following considerations are based on these pairings (and the sketched messenger sector) and do not explicitly depend on the discrete symmetries used to achieve them. In general these do not have to be chosen only to fulfill these pairings, but also to shape the flavon potential such that it gives the desired vacuum alignment (cf.\ \cite{Antusch:2010es, Antusch:2011sx}). Since this task will require different quantum numbers for the flavons $\phi_{2}$ and $\tilde{\phi}_{2}$, it might be desirable to introduce another Higgs representation $H_{24}$ or $\bar H_5$.
Due to this difference we also assume that $N_2$ is not paired with $\tilde{\phi}_2$.
However, these model building issues are beyond the scope of this paper.

\subsection{Resulting Yukawa Matrices and Right-handed Neutrino Masses}\label{Yukawa}
The Yukawa matrices after GUT and family symmetry breaking are defined in the following way:
\begin{align}
-\mathcal{L}_{Y} &= (Y_u^*)_{ij} Q_i u^c_j H_u + (Y^*_d)_{ij} Q_i d^c_j H_d + (Y^*_e)_{ij} L_i e^c_j H_d  \nonumber\\ 
                 &+ (Y^*_\nu)_{ij} L_i \nu^c_j H_u  + (M_{\nu^c})_{ij} \nu^c_i \nu^c_j + \mathrm{H.c.} \;.
\end{align}
This convention matches the one used by the Particle Data Group \cite{Amsler:2008zzb}. In our class of models the Yukawa matrices and $M_{\nu^c}$ have the following form at the GUT scale:
\begin{align}
Y_d &= \begin{pmatrix} 0 & \epsilon_{1} & - \epsilon_{1}  \\ \epsilon_{2} & \epsilon_{2} + \mathrm{i} \tilde{\epsilon}_{2} & \epsilon_{2} + w \tilde{\epsilon}_{2} \\ 0 & 0 & \epsilon_{3} \end{pmatrix} \;, \label{Eq:Yukd}\\
Y_e^T &= \begin{pmatrix} 0 & c_{1} \epsilon_{1} & - c_{1} \epsilon_{1} \\ c_{2} \epsilon_{2} & c_{2} \epsilon_{2} + \mathrm{i} \tilde{c}_{2} \tilde{\epsilon}_{2} & c_{2} \epsilon_{2} + w \tilde{c}_{2} \tilde{\epsilon}_{2} \\ 0 & 0 & c_3 \epsilon_{3} \end{pmatrix} \;, \label{Eq:Yuke} \\
Y_u &\equiv \begin{pmatrix} \hat{y}_u & 0 &  0 \\ 0 & \hat{y}_c& 0 \\ 0 &  0 & \hat{y}_t  \end{pmatrix} \;, \label{Eq:Yuku}\\
Y_\nu & \equiv \begin{pmatrix} 0 & \hat{y}_{\nu 2} \\ \hat{y}_{\nu 1} & \hat{y}_{\nu 2} \\ -\hat{y}_{\nu 1} & \hat{y}_{\nu 2} \end{pmatrix} \;, \label{Eq:Yuknu}\\
M_{\nu^c} & \equiv \begin{pmatrix} M_{\nu^c,1} & 0 \\ 0 & M_{\nu^c,2} \end{pmatrix} \;. \label{Eq:Mnu}
\end{align}
We remark that in the definition for the Yukawa matrices we have introduced a complex conjugation to match the PDG convention. This is the reason why $+\mathrm{i}$ instead of $-\mathrm{i}$ appears in the 2-2 elements of the down-type quark and charged lepton Yukawa matrices. 

The matrices $Y_d$ and $Y_e^T$ are expressed in terms of the ``expansion parameters'' $\epsilon_{1}$, $\epsilon_{2}$, $\tilde{\epsilon}_{2}$ and $\epsilon_{3}$, which are defined via the relations:
\begin{equation} \label{Eq:epsilons}
\begin{split}
\frac{v_{24}}{M_{1}^\prime} \frac{ \langle \phi_{1} \rangle }{M_{1}}  = \begin{pmatrix}  0 \\  1\\  - 1  \end{pmatrix} \epsilon_{1} \;, \quad
\frac{v_{24}}{M_{2}^\prime} \frac{ \langle \phi_{2} \rangle }{M_{2}}  &= \begin{pmatrix} 1 \\  1 \\ 1  \end{pmatrix}  \epsilon_{2} \;, \quad
\frac{v_{24}}{M_{3}^\prime} \frac{ \langle \phi_{3} \rangle }{M_{3}}  = \begin{pmatrix}  0 \\  0\\  1  \end{pmatrix} \epsilon_{3} \;, \\
\frac{v_{24}}{\tilde{M_{2}}^\prime} \frac{ \langle \tilde\phi_{2} \rangle }{\tilde M_{2}}  &= \begin{pmatrix}  0 \\  - \mathrm{i} \\  w  \end{pmatrix} \tilde{\epsilon}_{2} \;.
\end{split}
\end{equation}

The structures of $Y_d$ and $Y_e^T$, and its close relation, will lead to testable predictions of our class of models and the parameters $\epsilon_{1}$, $\epsilon_{2}$, $\tilde{\epsilon}_{2}$ and $\epsilon_{3}$ will be determined by a fit to the data for down-type quark and charged lepton masses and to the quark mixing parameters. The factors $c_1 = c_2 = c_{3} = -3/2$ and $\tilde c_{2} = 6$ are Clebsch--Gordan coefficients arising from GUT symmetry breaking and have been proposed recently in \cite{Antusch:2009gu} as new possible GUT relations in gravity mediated SUSY breaking scenarios. 
Clearly, the obviously incorrect $SU(5)$ relation $Y_d = Y_e^T$ would be obtained if all Clebsch--Gordan coefficients were one.  
To provide an idea of the size of the expansion parameters, let us say that in SUSY models with $\tan \beta \approx 30$ we will obtain small $\epsilon_{1}$, $\epsilon_{2}$ and $\tilde{\epsilon}_{2}$ of $\mathcal{O}(10^{-3})$ or less and $\epsilon_{3}$ of $\mathcal{O}(0.1)$. 

The parameters in the diagonal matrix $Y_u$ simply have a one-to-one correspondence to the up-type quark masses and the ratios of the squared elements of $Y_\nu$ over the right-handed neutrino masses in $M_{\nu^c}$ stands, via the seesaw mechanism, in direct relation to the observed light neutrino mass splittings. By the structure of the vevs of $\phi_{1}$ and $\phi_{2}$ the leptonic mixing matrix is of tri-bimaximal form to leading order and receives calculable corrections by the (small) mixings of $Y_e$, resulting in predictions for the neutrino mixing parameters as we will discuss later in the paper. 

Defining the ratios of flavon (and $H_{24}$) vevs over messenger masses as
\begin{equation}\label{Eq:ZetaDefinitions}
\zeta_{(i)} \equiv \frac{|\langle\phi_{(i)}\rangle|}{M_{(i)}} \quad \text{and} \quad \zeta^{24}_{(i)} \equiv \frac{v_{24}}{M^\prime_i}\;,
\end{equation} 
with $v_{24}$ being the vev of $H_{24}$, the entries of $Y_u$, $Y_\nu$ and $M_{\nu^c}$ are defined as
\begin{gather}
\hat{y}_u =  \frac{M_{1}^2}{M^u_{1}M^{\prime u}_{1}} \zeta_1^2\,,\quad 
\hat{y}_c = \frac{M_{2}^2}{M^u_{2}M^{\prime u}_{2}} \zeta_2^2\,,\quad  
\hat{y}_t = y_3 \,,\\
\hat{y}_{\nu 1} = \frac{1}{\sqrt{2}} \zeta_1 \,,\quad  
\hat{y}_{\nu 2} = \frac{1}{\sqrt{3}} \zeta_2 \,, \\
M_{\nu^c,1} =  \zeta_1^2 \frac{M_{1}^2}{M^{\prime u}_{1}}    \,,\quad
M_{\nu^c,2} =  \zeta_2^2 \frac{M_{2}^2}{M^{\prime u}_{1}}     \,.
\end{gather}

Now we have basically defined the building blocks of our class of flavour models. 
We would like to remark that within this class of models it is possible to define a specific version of a model by specifying the full set of (discrete) symmetries, the Higgs content at the GUT level, the vacuum alignment mechanism as well as the messenger sector of the theory. Note that, a priori, any family symmetry with real triplets can be used. 
Many predictions of the model class, however, are quite generic and will not depend on the realization in a specific model version.

\section{The SUSY Specific Parts} \label{sec:SUSY}
In principle a model belonging to the class discussed in the previous section may also be realised in a non-supersymmetric setup. 
However, if SUSY will be discovered, it will be necessary to extend any given model of flavour to a supersymmetric one and to perform a careful analysis of its predictions in the SUSY sector. The first task, to be discussed in this section, is to specify the superpotential $W$ and the K\"ahler potential $K$ of the SUSY version of the model, which is consistent with the symmetries of the model and can be generated by its messenger field content. From $W$ and $K$ we can calculate the canonically normalized Yukawa couplings  as well as the soft SUSY breaking sfermion mass matrices and the trilinear sfermion coupling matrices.

In the following, we will focus on the scenario that SUSY is broken in a hidden sector and then mediated to the visible sector by gravity. 
We will also assume complete {\it sequestering} of $K$ and $W$, meaning that we can split the superpotential and K\"ahler potential into two parts 
\begin{align*}
W &= W_{\rm vis} + W_{\rm hid} \;, \\
K &= K_{\rm vis} + K_{\rm hid} \;,
\end{align*}
respectively, where the visible superpotential is just $W_{\rm vis} = W_{Y_{d,l}} + W_{Y_u} + W_{Y_\nu} + W_{M_{\nu^c}} + \tilde W$ and $K_{\rm vis} = K_F + K_T + K_H + ...$, where the dots refer to terms for the other visible sector superfields, and where $\tilde W$ contains additional superpotential terms including, for instance, the flavon and GUT Higgs potentials. 

Furthermore, we will assume a gauge kinetic function such that unification of gaugino masses is realised at the GUT scale.

\subsection{The Superpotential}
The superpotential can be obtained by simply promoting the fields in the effective Lagrangian to chiral superfields. For our example class of models this gives\footnote{To simplify notation we will use the same letters for the fields also to denote the associated superfields.}
\begin{align}
W_{Y_{d,l}} &= \sqrt{2}\,F H_{24} \left(\frac{\phi_{1}}{M_{1}M_{1}^\prime} T_1 + \frac{\phi_{2}}{M_{2}M_{2}^\prime} T_2 +  \frac{\phi_{3}}{M_3 M_3^\prime} T_3 \right) \bar{H}_5 + \sqrt{2}\,
T_2 \frac{H_{24}}{\tilde{M}_2^\prime} \frac{\tilde{\phi}_{23}}{\tilde{M}_{2}} F \bar{H}_{5} \;, \label{Eq:WYl} \\
W_{Y_u} &= \frac{1}{4}\left( y_3 T_3^2 + \frac{\phi_{2}^2}{M^u_{2}M^{\prime u}_{2}} T_2^2 + \frac{\phi_{1}^2}{M^u_{1}M^{\prime u}_{1}} T_1^2  \right) H_5 \;, \label{Eq:WYu} \\
W_{Y_\nu} &= F \left(\frac{\phi_{1}}{M^{\prime u}_{1}} N_1 +  \frac{\phi_{2}}{M^{\prime u}_{2}} N_2 \right) H_5 \;, \label{Eq:WYnu} \\
W_{M_{\nu^c}} &= \frac{\phi_{1}^2}{M_{1}} N_1^2 + \frac{\phi_{2}^2}{M_{2}} N_2^2 \;. \label{Eq:WMR}
\end{align}
Promoting also the messenger fields to superfields, the effective operators in the superpotential can be generated by 
integrating out the heavy messenger superfields.  Integrating out these messengers also induces effective operators in the K\"ahler potential, as we discuss now.

\subsection{The K\"ahler Potential}

\begin{figure}
\centering
\includegraphics[scale=0.7]{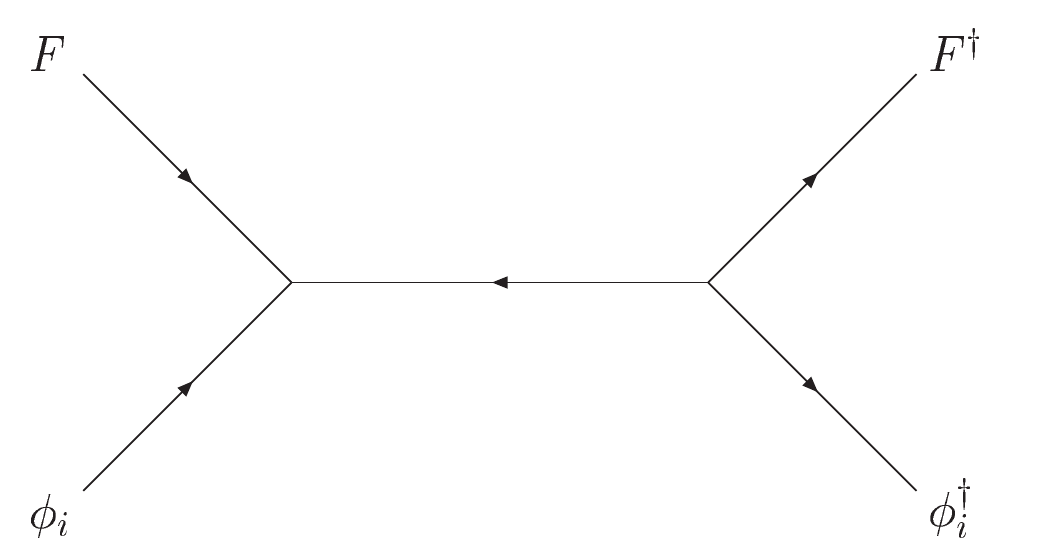}
\caption{Supergraph diagrams inducing the effective K\"ahler potential operators for the fiveplet superfield $F$. \label{Fig:messengers_KF}}
\end{figure}

Integrating out the messenger superfields generates not only the above superpotential operators but also effective operators in the K\"ahler potential. An example, within our class of models, is shown in Fig.~\ref{Fig:messengers_KF}. The messenger superfield is the same that is interchanged in the left side of the left diagram in Fig.~\ref{Fig:messenger_d}.
After family symmetry breaking, the kinetic terms are generically generated with an a priori non-canonical structure and have to be canonically normalized. This, in turn, has an effect on the fermion flavour structure, namely on the Yukawa couplings, as we will discuss in Sec.\ \ref{sec:normalisedYukawas}.

From the diagrams that generate the superpotential terms, using the argument from above, one can easily infer the non-canonical K\"ahler potential operators which are generated by the messenger fields.\footnote{We note that, in addition, there may also be effects from Planck scale suppressed operators. With the messenger field masses sufficiently below the Planck scale, these effects are subdominant and can be neglected.}
In our example class of models, the K\"ahler potential of the visible sector fields $T_i$ and $F$, including these operators, is given by
\begin{align}\label{Eq:EffectiveKaehlerT}
K_{T} =& T_1 T_1^\dagger k^{(T)}_1 + T_2 T_2^\dagger \left( k^{(T)}_2 + h^{(2)} \frac{ H_{24} H_{24}^\dagger}{\tilde{M}^{\prime\,2}_2} \right) + T_3 T_3^\dagger k^{(T)}_3 \;,\\
K_{F} =& F F^\dagger \left( k^{(F)} {\bf 1} + l^{(F)}_{1} \frac{ \phi_{1} \phi_{1}^\dagger}{M_{1}^2} + l^{(F)}_{2} \frac{ \phi_{2} \phi_{2}^\dagger }{M_{2}^2} + l^{(F)}_{3} \frac{ \phi_{3} \phi_{3}^\dagger }{M_3^2} \right) \;. \label{Eq:EffectiveKaehlerF}
\end{align}
Note that $K_{T}$ does not contain non-renormalizable operators with flavons, while such operators are present in $K_{F}$ and lead to non-diagonal terms after family symmetry breaking. In contrast, $K_F$ does not contain non-renormalizable operators\footnote{This is only true up to dimension 6. There are dimension 8 operators involving both flavons and $H_{24}$, which we, however, neglect.} with $H_{24}$. Gauge and family indices have been suppressed and are assumed to be contracted between the matter field and the flavon/GUT-Higgs field, see Fig.\ \ref{Fig:messengers_KF}. The couplings $l^{(X)}_i$ and $h^{(2)}$ depend on the fundamental couplings of the messengers, but are essentially additional free $\mathcal{O}(1)$ parameters as they do not appear somewhere else.

From the diagrams in Fig.\ \ref{Fig:messenger_d} and \ref{Fig:messenger_u} one can see that there is also a flavon-dependent contribution to the K\"ahler potential of the Higgs field $\bar{H}_5$:
\begin{equation} \label{Eq:EffectiveKaehlerH}
K_{\bar{H}_5} = \bar{H}_5 \bar{H}_5^\dag \left(k^{(H)} + \tilde{l}^{(H)}_{2} \frac{ \tilde{\phi}_{2} \tilde{\phi}_{2}^\dagger}{\tilde{M}_{2}^2}\right) \;.
\end{equation}

We have not given here the K\"ahler potential for the right-handed neutrino fields since its effects are negligible for phenomenology or can be absorbed into other unknown quantities. We have also dropped terms induced by the couplings of the $F$ and $T$ fields to $H_5$ and $\bar{H}_5$, since the resulting non-universal terms are suppressed by powers of the electroweak symmetry breaking (EWSB) vevs over the heavy messenger scale.

\subsection{Canonically Normalized Yukawa Matrices}\label{sec:normalisedYukawas}

As typical in SUSY/SUGRA flavour models, the kinetic terms in our example class of models are generically non-canonical and have to be canonically normalized, which has an impact on the fermion flavour structure. We will now discuss the leading order effects of this procedure.

The transformation to such a basis can be achieved as follows (see, e.g.\ \cite{Antusch:2007re}):
\begin{equation}
 F \to P_F^{-1} F\;,
\end{equation}
with
\begin{equation}
P^{-1\,\dag}_F \tilde{K}_{F} P^{-1}_F = {\bf 1}\,, ~{\text{ or equivalently }} ~~ \tilde{K}_{F} = P^{\dag}_F P_F\,,
\end{equation}
where $\tilde{K}_{F}$ is the K\"ahler metric defined as $ \tilde{K}_{F} \equiv \frac{\partial^2 K_F}{\partial F_i \partial F_j^\dag}$. Likewise, we define the transformation matrices $P_T$ and $P_{\bar{H}_5}$ which canonically normalize $K_T$ and $K_{\bar{H}_5}$. 

As a consequence of the field redefinitions, the Yukawa matrices $Y_{d,e}$ are transformed as follows:
\begin{equation}
Y_d \rightarrow (P_Q^{-1})^T Y_d P_{d^c}^{-1}\,,~~ Y_e \rightarrow (P_L^{-1})^T Y_e P_{e^c}^{-1} \,.
\end{equation}
The rescaling factor needed for canonically normalizing $K_{\bar{H}_5}$ is readily absorbed into the definition of the $\epsilon$'s. In contrast to $Y_{d,e}$, the Yukawa matrix $Y_u$ remains unchanged in form as one can absorb all of $P_T$ in a redefinition of $\hat{y}_u$, $\hat{y}_c$, $\hat{y}_t$.

Several remarks are in order here:
\begin{itemize}

\item The parameters which are directly related to the fermion masses and mixings are the $\epsilon$'s. On the other hand, in our example class of models the $\epsilon$'s depend on the product of two ratios of flavon vevs over messenger masses, for example $\epsilon_3 \sim \zeta_{(3)} \zeta^{24}_{(3)}$, and similarly for the other $\epsilon$'s as defined in the previous section. In the following, we will assume the case that $\zeta^{24}_{(2)},\zeta^{24}_{(1)},$ and $\zeta_{(\tilde{2})}$ are comparatively large ($\gtrsim\mathcal{O}(0.1)$). This in turn implies that, as we will confirm later by the fit to the data, that $\zeta_{(2)}$, $\zeta_{(1)}$, $\zeta^{24}_{(\tilde{2})}$ are very small such that we can neglect the corresponding canonical normalization effects in a leading order analysis. The only $\epsilon$ parameter which will not turn out to be very small is  $\epsilon_3$, and therefore we will introduce both parameters $\zeta_{(3)}$ and $ \zeta^{24}_{(3)}$ explicitly.

\item Since $K_T$ also involves couplings to the GUT breaking vev of $H_{24}$ but $K_{F}$ does not, we get the relations
$P_F = P_{d^c} = P_{L}$ and $P_Q \ne P_{u^c} \ne P_{e^c}$. This modifies the CG factors between the three fields contained in the tenplet $T_2$, i.e. $c_{2}$ and $\tilde{c}_{2}$, due to the vev of $H_{24}$ which treats those fields differently. However, since $\zeta^{24}_{(\tilde{2})}$ is very small, as stated above, this can be neglected. 

\end{itemize}

Under the assumptions specified above the main effect of canonical normalization stems from the change in $Y_d$ and $Y_e$ due to rescaling of the third component of the field $F$, which is parameterised by $\zeta_3$. With the definition $\hat{\zeta}_3 = \zeta_3 ({l^{(F)}_3}/{k^{(F)}})^{1/2}$, and after absorbing all other rescalings in the $\epsilon$'s, the Yukawa matrices have the following form at the GUT scale to leading order in $\hat{\zeta}_3$:
\begin{align}
Y_u &= \begin{pmatrix} \hat{y}_u & 0 &  0 \\ 0 & \hat{y}_c& 0 \\ 0 &  0 & \hat{y}_t  \end{pmatrix} \;, \label{Eq:Yuku_can}\\
Y_d &= \begin{pmatrix} 0 & \epsilon_{1} & - \epsilon_{1} (1 - \frac{1}{2} \hat{\zeta}^2_3) \\ \epsilon_{2} & \epsilon_{2} + \mathrm{i} \, \tilde{\epsilon}_{2} & (\epsilon_{2} + w \tilde{\epsilon}_{2})(1-\frac{1}{2} \hat{\zeta}^2_3) \\ 0 & 0 & \epsilon_{3} (1-\frac{1}{2} \hat{\zeta}^2_3) \end{pmatrix} \;, \label{Eq:Yukd_can}\\
Y_e^T &= \begin{pmatrix} 0 & c_{1} \epsilon_{1} & - c_{1} \epsilon_{1} (1-\frac{1}{2} \hat{\zeta}^2_3) \\ c_{2} \epsilon_{2} & c_{2} \epsilon_{2} + \mathrm{i} \, \tilde{c}_{2} \tilde{\epsilon}_{2} & (c_{2} \epsilon_{2} + w \tilde{c}_{2} \tilde{\epsilon}_{2}) (1-\frac{1}{2} \hat{\zeta}^2_3) \\ 0 & 0 & c_3 \epsilon_{3} (1-\frac{1}{2} \hat{\zeta}^2_3) \end{pmatrix} \label{Eq:Yuke_can} \;, \\
Y_\nu &= \begin{pmatrix} 0 & \hat{y}_{\nu 2} \\ \hat{y}_{\nu 1} & \hat{y}_{\nu 2} \\ -\hat{y}_{\nu 1} (1-\frac{1}{2} \hat{\zeta}^2_3) & \hat{y}_{\nu 2} (1-\frac{1}{2} \hat{\zeta}^2_3) \end{pmatrix} \;. \label{Eq:Yuknu_can}
\end{align}

\subsection{Soft Breaking Terms in Supergravity} \label{sec:modelsoftterms}

Let us now turn to the soft SUSY breaking terms which can be calculated from a given $K$ and $W$. Please keep in mind that we restrict ourselves to the case of hidden sector SUSY breaking transmitted to the visible sector by gravity and that we consider for definiteness complete {\it sequestering} of $K$ and $W$ into hidden sector and visible sector parts. 
The hidden sector is assumed to contain chiral superfields which break SUSY by obtaining both a vev and a non-vanishing F-term of order $M_{\text{Pl}}$ and $m_{3/2} M_{\text{Pl}}$, respectively.

In supergravity before canonical normalization, the soft masses can be expressed in terms of the full visible sector K\"ahler metric $\tilde{K}$ \cite{Kaneetal}:
\begin{equation}
\tilde{m}^2_{\bar{a}b} = m_{3/2}^2 \tilde{K}_{\bar{a}b} - F_{\bar{m}} (\partial_{\bar{m}} \partial_{n} \tilde{K}_{\bar{a}b}) F_n
+ F_{\bar{m}} (\partial_{\bar{m}} \tilde{K}_{\bar{a}c} ) \tilde{K}^{-1}_{c\bar{d}} (\partial_{n} \tilde{K}_{\bar{d}b}) F_n \,,
\label{Eq:SUGRAsoftmass}
\end{equation}
where $n$($\bar{m}$) runs over all (conjugated) superfields while $b$($\bar{a}$) only runs over matter superfields. Clearly, given the sequestering assumption, only {\it visible sector} superfields whose F-term is not vanishing contribute to the second and third term of Eq.~\eqref{Eq:SUGRAsoftmass}.
In our class of models this is the case for the flavons and the GUT Higgs field, which acquire an F-term proportional to their vev \cite{Ross-Vives}:
\begin{equation}
F_{\Phi} = C_{\Phi} m_{3/2} \langle \Phi \rangle \,,
\label{Eq:fterm}
\end{equation}
where the coefficients $C_{\phi_{(i)}}$, $C_{H_{24}}$ could be computed after fully specifying the flavon and GUT Higgs superpotential.
Given the ``irreducible'' contribution $m_{3/2} \langle \Phi \rangle$ to $F_{\Phi}$ \cite{Ross-Vives},
we can assume the $C_{\Phi}$ to be real $\mathcal{O}(1)$ parameters\footnote{The specific conditions for this have been examined in \cite{AKMR} and \cite{Feruglio:2009iu}.}.

On the other hand the SUSY breaking trilinear terms \cite{Kaneetal} take the form:
\begin{equation}
A_{abc} = \text{e}^{\frac{K_{\rm hid}}{2 M_{\text{Pl}}^2}} \frac{W^*_{\rm hid}}{|W_{\rm hid}|} F_m \left(\left(\partial_m \frac{K_{\rm hid}}{M^2_{\text{Pl}}}\right) \partial_a \partial_b \partial_c W_{\rm vis} + \partial_m \partial_a \partial_b \partial_c W_{\rm vis} + \cdots \right)\,,
\label{Eq:SUGRAaterm_unrescaled}
\end{equation}
where subdominant terms have been omitted. However after SUSY breaking in the hidden sector also the Yukawa couplings appearing in $W_{\rm vis}$ are no longer the physical Yukawa couplings for the low energy theory but take the form
\begin{equation}
Y_{abc} = \text{e}^{\frac{K_{\rm hid}}{2 M_{\text{Pl}}^2}} \frac{W^*_{\rm hid}}{|W_{\rm hid}|} \partial_a \partial_b \partial_c W_{\rm vis} \;,
\label{Eq:SUGRAYukawas}
\end{equation}
multiplied by the same factor also multiplying $A_{abc}$.
This means that Eq.~(\ref{Eq:SUGRAaterm_unrescaled}) is simplified to
\begin{equation}
A_{abc} = F_h \left(\partial_h \frac{K_{\rm hid}}{M^2_{\text{Pl}}}\right) Y_{abc} + F_m \partial_m Y_{abc} + \cdots \,,
\label{Eq:SUGRAaterm}
\end{equation}
where $h$ runs over hidden sector fields and $m$ over visible sector fields. For this we used the fact that the rescaling factor only depends on hidden sector fields and thus commutes with $\partial_m$ due to sequestering.
We can then define the usual mSUGRA parameter $A_0 = F_h \partial_h \frac{K_{\rm hid}}{M^2_{\text{Pl}}}$ which depends on the unspecified hidden sector superpotential and K\"ahler potential. Regarding the Yukawa matrices we can absorb the factor\footnote{If we assume one hidden sector field $X$ with $F_X^2 = 3 m_{3/2}^2 M_{\text{Pl}}^2$ and canonical hidden K\"ahler potential we get as factor $\text{exp} ( {\frac{K_{\rm hid}}{2 M_{\text{Pl}}^2}} ) = \text{exp} ({\frac{A_0^2}{6 m_0^2}})$ ($\simeq 2$ if $A_0=2 m_0$).} $\text{exp} ({\frac{K_{\rm hid}}{2 M_{\text{Pl}}^2}}) W^*_{\rm hid} / |W_{\rm hid}|$ into the previous definition of our Yukawa matrices. However, the expressions linking the parameters $\epsilon$'s and $\zeta$'s would be then modified by a $\mathcal{O}(1)$ coefficient.

From these formulae it is apparent why sequestering simplifies our construction for two reasons:
\begin{itemize}
\item It is known that models of the considered class with family symmetries with real triplets do not protect all particle species against large flavour violation in their soft mass matrices (see, e.g.\ \cite{Antusch:2007re}). In our case, while the five-dimensional representation is protected, the ten-dimensional remains unprotected.
In particular, without sequestering there could be
Planck scale suppressed couplings to hidden fields on the diagonal of the
tenplet K\"ahler metric, inducing
$\mathcal{O}(1)$ splittings of the tenplet soft masses.
This would generate, in the basis where $Y_d$ is diagonal (the SCKM basis), 1-2 entries in
the soft mass matrices of the tenplet constituents which are only
Cabibbo suppressed.

\item With a sequestered K\"ahler potential all flavour violation effects come from the F-terms of the family symmetry breaking Higgs fields, i.e.\ the flavons, and in principle also from the F-terms of the GUT-Higgs fields. This clearly reduces the number of variables entering the soft terms and makes the model more predictive.
\end{itemize}

\subsubsection{Soft Masses}

From the expressions in Eqs.\ \eqref{Eq:EffectiveKaehlerT}, \eqref{Eq:EffectiveKaehlerF}, and \eqref{Eq:EffectiveKaehlerH}, we can now derive the structure of the soft-mass matrices at the GUT scale.

After canonical normalization the first term of Eq.~\eqref{Eq:SUGRAsoftmass} gives rise to a universal contribution to the soft masses. However, due to the coefficients $C_{\Phi}$, $F_{\bar{m}} (\partial_{\bar{m}} \partial_{n} \tilde{K}) F_n$ is not rotated away by the
same transformation making $\tilde{K}\rightarrow {\bf 1}$. Thus, such a term gives rise to non-universal soft masses already at the
GUT scale. The third term has the same effect, but provides only contributions $\propto \zeta^4$, which we neglect.

If we take $m_0 \equiv m_{3/2}$, in the canonical basis, the soft mass matrices for the matter fields then result  up to second order in $\hat{\zeta}_3$ (neglecting all the small $\zeta$'s):
\begin{align}
m^2_{\tilde{Q}} &= m_0^2 {\bf 1} \;, \\
m^2_{\tilde{U}} &= m_0^2  {\bf 1} \;, \\
m^2_{\tilde{D}} &= m_0^2   {\bf 1} - (C_{\phi_3})^2 \hat{\zeta}_{3}^2 m_0^2 \begin{pmatrix} 0 & 0 & 0 \\ 0 & 0 & 0 \\ 0 & 0 & 1 \\ \end{pmatrix} \;, \label{Eq:m2d}
\end{align}
and for the leptons
\begin{align}
m^2_{\tilde{L}} &= m_0^2  {\bf 1} - (C_{\phi_3})^2 \hat{\zeta}_{3}^2 m_0^2 \begin{pmatrix} 0 & 0 & 0 \\ 0 & 0 & 0 \\ 0 & 0 & 1\\ \end{pmatrix} \;, \label{Eq:m2l}\\
m^2_{\tilde{E}} &= m_0^2  {\bf 1} \;. 
\end{align}

Additionally the Higgs boson $\bar{H}_5$ which contains the light Higgs doublet $H_d$ gets a non-universal soft squared mass
\begin{equation}
m^2_{H_d} = (1 - (C_{\tilde{\phi}_{2}})^2 \hat{\zeta}_{\tilde{2}}^2) m_0^2 \;.
\label{Eq:m2Hd}
\end{equation}

If we assume $\hat{\zeta}_{\tilde{2}}$ to be $\mathcal{O}(0.1)$ then the non-universality amounts to only a few percent, which is smaller than the accuracy of the 1-loop RGEs for the soft terms we use in our analysis.

It is interesting to note here that due to the assumed pairing between different fields the soft mass terms are actually only diagonal for the tenplets. The soft terms for the fiveplets have off-diagonal terms which are small since they are generated by terms of order the small $\zeta^2$. The large $\hat{\zeta}_3$ does not generate off-diagonal terms. However, this is only the case in the flavour eigenbasis. Upon rotating to the basis where the Yukawa matrices are diagonal (the SCKM basis) these non-universal diagonal entries induce off-diagonal entries suppressed by mixing angles.

\subsubsection{Soft Trilinear Couplings}

The first term of Eq.~(\ref{Eq:SUGRAaterm}) does not give rise to flavour mixing, since it is proportional to the Yukawa matrix and hence diagonal in the same basis.

The term proportional to $\partial_m Y$ has the same flavour structure as the corresponding Yukawa matrix, but it is not necessarily proportional
to it, due to the flavon F-term coefficients $C_{\phi_{(i)}}$. Thus, the A-term matrices $A_d$ and $A_e$ will not be diagonal in the SCKM basis. As we will see, this represents the dominant source of flavour violation in our class of models. We will refer to this scheme as ``Trilinear Dominance''.

After canonical normalization, the trilinear terms before going to the SCKM basis look like:
\begin{align}
A_d =& \, A_0 Y_d + C_{H_{24}} m_0 Y_d  \nonumber\\
&+ m_0 \begin{pmatrix}
0 & C_{\phi_{1}} \epsilon_{1} & - C_{\phi_{1}} \epsilon_{1} (1-\frac{1}{2} \hat{\zeta}^2_3) \\
C_{\phi_{2}} \epsilon_{2} & C_{\phi_{2}} \epsilon_{2} + \mathrm{i} C_{\tilde{\phi}_{2}} \tilde{\epsilon}_{2} &
(C_{\phi_{2}} \epsilon_{2} + w C_{\tilde{\phi}_{2}} \tilde{\epsilon}_{2}) (1-\frac{1}{2} \hat{\zeta}^2_3) \\
0 & 0 &  C_{\phi_{3}} \epsilon_{3} (1-\frac{1}{2} \hat{\zeta}^2_3)
\end{pmatrix} \;,
\label{Eq:Ad}\\
A_e^T =& \, A_0 Y_e^T + C_{H_{24}} m_0 Y_e^T  \nonumber\\
&+ m_0 \begin{pmatrix}
0 & -\frac{3}{2} C_{\phi_{1}} \epsilon_{1} & \frac{3}{2} C_{\phi_{1}} \epsilon_{1} (1-\frac{1}{2} \hat{\zeta}^2_3) \\
-\frac{3}{2} C_{\phi_{2}} \epsilon_{2} & -\frac{3}{2} C_{\phi_{2}} \epsilon_{2} + 6 \mathrm{i} C_{\tilde{\phi}_{2}} \tilde{\epsilon}_{2} &
(-\frac{3}{2} C_{\phi_{2}} \epsilon_{2} + 6 w C_{\tilde{\phi}_{2}} \tilde{\epsilon}_{2}) (1-\frac{1}{2} \hat{\zeta}^2_3) \\
0 & 0 & -\frac{3}{2} C_{\phi_{3}} \epsilon_{3} (1-\frac{1}{2} \hat{\zeta}^2_3)
\end{pmatrix} \;,
\label{Eq:Ae}\\
A_u =& \, A_0 Y_u + m_0 \begin{pmatrix}      
2 C_{\phi_{1}} \hat{y}_u & 0 & 0 \\
0 & 2 C_{\phi_{2}} \hat{y}_c &
0 \\
0 & 0 & 0
\end{pmatrix} \;,
\label{Eq:Au}\\
A_\nu =& \, A_0 Y_\nu + m_0 \begin{pmatrix} 0 & C_{\phi_{2}} \hat{y}_{\nu 2} \\ C_{\phi_{1}} \hat{y}_{\nu 1} & C_{\phi_{2}} \hat{y}_{\nu 2} \\ - C_{\phi_{1}} \hat{y}_{\nu 1} (1-\frac{1}{2} \hat{\zeta}^2_3) & C_{\phi_{2}} \hat{y}_{\nu 2} (1-\frac{1}{2} \hat{\zeta}^2_3) \end{pmatrix} \;.
\end{align}
Note that the contribution proportional to $C_{H_{24}}$ is just $Y_{d,e}$ times a factor and thus does not give any additional flavour violating effects, but only a splitting between $A_{0,u}$ and $A_{0,d/e}$. Thus for simplification, we neglect these terms and assume 
\begin{equation}
C_{H_{24}} = 0
\end{equation}
for the rest of our analysis.
The 3-3 element of the corrections to $A_u$ vanishes since we assume that for the third generation the messengers are absent as the top Yukawa coupling is generated on tree-level.
$A_\nu$ is of less phenomenological importance as the right-handed sneutrinos get integrated out not much below the GUT scale.

\subsection{SUSY Threshold Corrections} \label{sec:SUSYThresholdCorrections}

So far we have discussed the impact of a flavour model on the structure of the soft SUSY breaking sector at very high energy scales. However, to fit a SUSY flavour model to the available data, there is a certain class of supersymmetric corrections at the low scale which is mandatory to include, in particular in the large (or moderate) $\tan \beta$ region. These are the well known SUSY threshold corrections \cite{SUSYthresholds}.

For simplicity, in the semi-analytic part we will use the formulae for the $\tan \beta$-enhanced SUSY threshold corrections in the EW unbroken phase, where effects of electroweak symmetry breaking are neglected (see, e.g.\ \cite{Antusch:2008tf}). We note, however, that it is important to include the SUSY threshold corrections for all three families of down-type quarks and charged leptons. The full formulae including EW symmetry breaking effects are given, e.g.\ in \cite{Pierce:1996zz}. 

One-loop matching between the SM quantities and the MSSM ones at the SUSY scale $M_{\text{SUSY}}$ leads to the approximate relations
\begin{align}
   y^{\text{SM}}_{e,\mu,\tau} &= (1 + \epsilon_l \tan\beta)~ y^{\text{MSSM}}_{e,\mu,\tau} \cos \beta \;,\\
   y^{\text{SM}}_{d,s} &= (1 + \epsilon_q \tan\beta)~ y^{\text{MSSM}}_{d,s} \cos \beta \;,\\
   y^{\text{SM}}_{b} &= (1 + (\epsilon_q + \epsilon_A) \tan\beta)~ y^{\text{MSSM}}_b \cos \beta \;,
\end{align}
for the Yukawa couplings and  
\begin{align}
   \theta^{\text{SM}}_{i3} &= \frac{1 + \epsilon_q \tan\beta}{1 + (\epsilon_q + \epsilon_A) \tan\beta}~ \theta^{\text{MSSM}}_{i3} \;, \\
   \theta^{\text{SM}}_{12} &= \theta^{\text{MSSM}}_{12} \;, \\
   \delta^{\text{SM}}_{\text{CKM}} &= \delta^{\text{MSSM}}_{\text{CKM}} \;,
\end{align}
for the quark mixing parameters. 
The one-loop SUSY corrections are parameterised in terms of the new $\epsilon$ parameters. We see that although the $\epsilon$ are loop suppressed the corrections can be sizeable if $\tan \beta$ is large enough. For explicit formulae and typical numerical values for the loop corrections, see \cite{Antusch:2008tf}, where we find
\begin{align}
\epsilon_l \tan\beta &= (\epsilon_e^{W} + \epsilon_e^{B}) \tan\beta = - (0.072 \div 0.020) \frac{\tan\beta}{30} \;, \label{eq:rangeLeptonThrCorr}\\
\epsilon_q \tan\beta &= (\epsilon_d^{G} + \epsilon_d^{W} + \epsilon_d^{B}) \tan\beta = (0.030 \div 0.248) \frac{\tan\beta}{30}\;, \label{eq:quarkLeptonThrCorr} \\
\epsilon_A \tan\beta &= \epsilon^y \tan\beta = \sgn A_t (0.025 \div 0.140) \frac{\tan\beta}{30}\;. \label{eq:rangeBottomYukawaThrCorr}
\end{align}
We should note that -- apart from neglecting effects of electroweak symmetry breaking and sfermion mixing -- this simplified parameterisation assumes that the corrections due to supersymmetric gauge interactions are universal for all three generations. In general this is not quite correct since the sfermion masses entering the formulae are in general not degenerate. Even in the CMSSM where the masses are set to be equal at the GUT scale at low energies the third generation sfermions are lighter due to the effect of large Yukawa couplings on the running. 
In the full numerical analysis of Sec.\ \ref{sec:num}, this will be taken into account.

We will also see that these corrections enable us to constrain the CMSSM parameter space if we match our class of models to the experimental data at low energies.

\subsection{Flavour and CP Violating Effects}
\label{sec:flavor-effects}

The first thing to note is that in our class of models the leading order structure of the soft SUSY breaking scalar masses 
at the GUT scale is close to the one of the CMSSM. 

A first potential deviation from the CMSSM flavour structure is represented by the non-universality of the third generation 
RH down squarks and LH sleptons given by the terms proportional to $(C_{\phi_3})^2 \hat{\zeta}_{3}^2$ in Eqs.~(\ref{Eq:m2d}, \ref{Eq:m2l}).
This could in principle determine a misalignment of the quark (lepton) and squark (slepton) mass matrices already at the GUT scale. In fact, 
in the basis where the corresponding Yukawa matrices are diagonal (the so-called SCKM basis), $m^2_{\tilde{D}}$ and $m^2_{\tilde{L}}$ 
will acquire off-diagonal entries in the 1-3 and 2-3 sectors, approximately given by the product of the sfermion mass splitting 
and the corresponding rotation to the SCKM basis. For instance, we have:
\begin{equation}
(m^2_{\tilde{D}})_{i\neq 3}^{\rm SCKM}=(V^d_R m^2_{\tilde{D}} V^{d\dag}_R)_{i\neq 3} \simeq 
-m^2_0 (C_{\phi_3} \hat{\zeta}_{3})^2 (V^d_R)_{i 3}\,,
\label{Eq:non-univ}
\end{equation}
where $V^d_R$ is one of the two unitary matrices diagonalizing $Y_d$:
\begin{equation}
\hat{Y}_d = V^{d\dag}_L Y_d V^{d}_R\,.
\end{equation}
However, we see from Eq.~(\ref{Eq:Yukd_can}) that the RH 1-3 and 2-3 rotations in $V^{d}_R$ are much smaller 
than the LH ones in $V^{d}_L$
(which accounts for the CKM angles). Indeed, they even vanish in the leading order approximation $(V^d_R)_{i3}\sim (Y_d)_{i3}/(Y_d)_{33}$. 
As a consequence, the source of flavour violation given by Eq.~(\ref{Eq:non-univ}) should
result subdominant with respect to the effects we discuss below. In the leptonic sector, it will turn out from the 
diagonalization of $Y_e$ (with numbers for the $\epsilon$'s as given in Tab.\ \ref{tab:numericalvaluesformodelparameters}) 
that $(V^e_R)_{23}\sim10^{-2}$. This can provide a source of $\tau$-$\mu$ lepton flavour violation 
at the level we discuss below. As we will see, such an effect is, however, of little phenomenological relevance, 
since $\tau$-$\mu$ transitions result hardly in the reach of future experiments.

Most notably, at the GUT scale the most relevant source of flavour violation is therefore represented by the trilinear couplings, 
$A_d$ and $A_e$, Eqs.~(\ref{Eq:Ad}, \ref{Eq:Ae}). We will refer to this scheme as ``Trilinear Dominance''. The trilinear coupling 
matrices have a flavour structure which is similar to that of the corresponding Yukawa matrices,
but they are not in general aligned to the Yukawas in flavour space. Such misalignment, 
as previously observed in the context of $SU(3)$ \cite{Ross-Vives, AKMR, su3edms} and $A_4$ \cite{Feruglio:2009iu, Feruglio:2009hu} 
flavour models, is a consequence of the flavon F-term coefficients $C_{\phi_{(i)}}$, 
and it will clearly determine that the A-term matrices are in general not diagonal in the SCKM basis.
Therefore, barring the unnatural case of all $C_{\phi_{(i)}}$ acquiring a common value,
the A-terms will be a source of potentially large flavour violating effects in our class of models.\footnote{On the other hand, we remark that depending on the flavon Higgs potential, the $C_{\phi_{(i)}}$ can also be strongly suppressed \cite{AKMR}. A discovery of SUSY particles at the LHC, but non-observation of any flavour violation effects, may thus point to a vacuum alignment with suppressed $C_{\phi_{(i)}}$, e.g.\ of the type discussed in \cite{AKMR}.} 
Effects of large trilinear terms have also been studied in \cite{Crivellin:2008mq, Crivellin:2009sd}.

Moreover, the large A-term matrices will induce flavour violating off-diagonal entries in the scalar mass matrices as well, through the RG evolution of these
parameters from the GUT scale to low energy. This latter effect can be understood by means of approximate leading-log
solutions of the 1-loop MSSM RGEs. Working in the basis of diagonal down quark and lepton Yukawas and neglecting for the moment 
the contribution of Eq.~(\ref{Eq:non-univ}), we have for the squark sector:
\begin{align}
(m^2_{\tilde{Q}})_{i\neq j} &\simeq (m^2_{\tilde{Q}})^{\rm CMSSM}_{i\neq j} - 
\frac{1}{8 \pi^2} (A_d A_d^\dag)_{ij} \ln \left(\frac{M_{\rm GUT}}{M_{\rm SUSY}}\right) \,,\label{Eq:m2Q-ll}\\
(m^2_{\tilde{D}})_{i\neq j} &\simeq - \frac{1}{4 \pi^2} (A_d^\dag A_d)_{ij} \ln \left(\frac{M_{\rm GUT}}{M_{\rm SUSY}}\right)\,,
\label{Eq:m2D-ll}
\end{align}
where $(m^2_{\tilde{Q}})^{\rm CMSSM}_{i\neq j}$ represents the usual CMSSM contribution, proportional to entries of the CKM matrix. Keep in mind, that in the basis we have chosen there are only off-diagonal elements generated in $m^2_{\tilde{Q}}$, within the CMSSM.
Analogously for the sleptons, we have:
\begin{align}
(m^2_{\tilde{L}})_{i\neq j} &\simeq - \frac{1}{8 \pi^2} (A_e A_e^\dag)_{ij} \ln \left(\frac{M_{\rm GUT}}{M_{\rm SUSY}}\right)\,,
\label{Eq:m2L-ll}\\
(m^2_{\tilde{E}})_{i\neq j} &\simeq - \frac{1}{4 \pi^2} (A_e^\dag A_e)_{ij} \ln \left(\frac{M_{\rm GUT}}{M_{\rm SUSY}}\right)\,,
\label{Eq:m2E-ll}
\end{align}
where the subdominant SUSY see-saw contribution to $(m^2_{\tilde{L}})_{i\neq j}$ has been omitted.
From the expressions above, we can see that at low energy we have sources of flavour violation both in the LH and in the RH scalar sectors,
as well as in the left-right mixing sector (the A-terms themselves). What is interesting is that all these sources of
flavour violation are determined by the matrices $A_d$ and $A_e$, i.e. they are controlled by the limited set of free parameters
$C_{\phi_{(i)}}$ (while the other parameters, such as the $\epsilon_{(i)}$, are determined by the fit of the fermion masses and mixings), 
and the effects in the squark and slepton sectors are linked by the underlying $SU(5)$ GUT. 

In order to give an estimate of the above discussed effects, we provide the following approximate expression of 
$A_d$ in the SCKM basis (the expression for $A^T_e$ is analogous with the substitution $\epsilon_i \to c_i \epsilon_i$):
\begin{align}
& A^{\rm SCKM}_d \simeq \, A_0 \hat{Y}_d +  m_0 \times \nonumber\\
& \begin{pmatrix}
\left(C_{\phi_1}+C_{2\tilde{2}}-\mathrm{i}C_{2\tilde{2}}\frac{\epsilon_{2}}{\tilde{\epsilon}_{2}}\right)\frac{\epsilon_{1} \epsilon_{2}}{\tilde{\epsilon}_{2}} &
\left(\mathrm{i}C_{2\tilde{2}}\frac{\epsilon_{2}}{\tilde{\epsilon}_{2}} -C_{1\tilde{2}}\right)\epsilon_{1}  & 
\left( \mathrm{i}(C_{13}+C_{\tilde{2}3}w\frac{\epsilon_{2}}{\tilde{\epsilon}_{2}}) -C_{\tilde{2}3}w\right)\epsilon_1 
(1-\frac{1}{2} \hat{\zeta}^2_3)\\
C_{2\tilde{2}} \epsilon_{2}-\mathrm{i}C_{2\tilde{2}} \frac{\epsilon_{2}^2}{\tilde{\epsilon}_{2}} & 
C_{\tilde{\phi}_{2}} \tilde{\epsilon}_{2} + \mathrm{i}C_{2\tilde{2}} \epsilon_{2}&
\left(C_{23}\epsilon_{2} + w C_{\tilde{2}3} \tilde{\epsilon}_{2}\right) (1-\frac{1}{2} \hat{\zeta}^2_3) \\
\mathcal{O}\left(\frac{\epsilon_2 \tilde{\epsilon}_2 }{\epsilon_3}\right) & 
\mathcal{O}\left(\frac{\tilde{\epsilon}_2^2 }{\epsilon_3}\right) &  
C_{\phi_{3}} \epsilon_{3} (1-\frac{1}{2} \hat{\zeta}^2_3)
\end{pmatrix},
\label{Eq:Adsckm}
\end{align}
where $\hat{Y}_d$ is the diagonalized Yukawa matrix and
$C_{ij}=C_{\phi_{(i)}}-C_{\phi_{(j)}}$.

From the expression above, we can see that, as expected, all the flavour-violating 
entries would vanish if $Y_d$ and $A_d$ were aligned (i.e. $C_{\phi_{1}}=C_{\phi_{2}}=C_{\tilde{\phi}_{2}}=C_{\phi_{3}}$).
Let us also notice that $A^{\rm SCKM}_d$ and $A^{\rm SCKM}_e$ acquire complex entries both on the diagonal and off-diagonal. This 
implies both flavour conserving and flavour violating contributions to the EDMs. 

We can now estimate the size of the flavour violating mass insertions (MIs), as usual defined as:
\begin{equation}
(\delta^f_{\rm XY})_{i\neq j} = \frac{(\Delta^{\tilde{f}}_{\rm XY})_{i\neq j} }{\bar{m}^2_{\tilde{f}}}\,,~~~({\rm X,\,Y=L,\,R})
\label{Eq:MIs}
\end{equation}
where $\bar{m}^2_{\tilde{f}}$ is the average sfermion squared mass and $\Delta^{\tilde{f}}_{\rm XY}$ are $3\times3$ blocks of the 
low-energy sfermion mass matrix. The off-diagonal flavour violating entries of $\Delta^{\tilde{f}}_{\rm XY}$ read:
\begin{equation}
(\Delta^{\tilde{f}}_{\rm LL,\,RR})_{i\neq j}  = (m^2_{\tilde{f}_{L,\,R}})_{i\neq j}\,, \quad
(\Delta^{\tilde{f}}_{\rm LR})_{i\neq j} = v_{d} (A^\dag_f)_{i\neq j} ~~
\left(\Delta^{\tilde{f}}_{\rm RL} = \Delta^{\tilde{f}\,\dagger}_{\rm LR}\right)   \,, 
\end{equation}
where $v_d$ is the down-type Higgs vev, $v_d\equiv v \cos\beta$ (we consider in fact $f=\ell,\,d$).
In the SCKM basis, the parameters $(\delta^f_{\rm XY})_{i\neq j}$ provide a measurement of the misalignment among fermions 
and sfermions, responsible for the flavour changing processes.

Making use of Eqs.~(\ref{Eq:m2Q-ll}-\ref{Eq:m2E-ll}) together with Eq.~(\ref{Eq:Adsckm}), it is possible to obtain good estimates
of the mass insertion parameters $(\delta^f_{\rm XY})_{ij}$, including the CP violating phases. The results can be then compared
with the bounds on mass insertions provided in the literature \cite{Altmannshofer:2009ne, Isidori:2010kg, sleptonarium, paride}. 
Such estimates also give a useful indication of the channels where the largest deviations from the SM predictions are 
expected and can be used to cross-check the results of the full numerical analysis (which will include the resolution of the 
full 1-loop RGEs, instead of using the leading-log approximation) presented in Sec.\ \ref{sec:numerical}. 
Some relevant examples will be provided in Sec.\ \ref{sec:estimate-fcnc}.

\section{Semi-analytical Analysis}

After having discussed the GUT scale structure of the Yukawa matrices and the soft SUSY breaking terms we will now start with the discussion of low energy predictions of the considered class of models, first in a semi-analytical approach and then we will also present numerical results in the next section.

We will start our discussion with some analytical estimates for the masses and mixing parameters at the GUT scale in the quark and the lepton sector which already show that our class of models is in principle realistic. Then we will also discuss the impact of the SUSY threshold corrections. Especially, we will discuss how these corrections can be used to constrain the CMSSM parameter space in our class of models.

\subsection{The Quark Sector}

We start with the discussion of the quark sector.
Since the up quark Yukawa matrix is diagonal we can directly determine the CKM mixing parameters from the down quark Yukawa matrix. We find the following approximate relations at the GUT scale:
\begin{align}
   \theta^{\text{CKM}}_{12}  &=  \left\vert \frac{\epsilon_1}{\epsilon_2 - \ci \, \tilde{\epsilon}_2} \right\vert \left( 1 - \left\vert\frac{\epsilon_2}{\epsilon_2 - \ci \, \tilde{\epsilon}_2}\right\vert^2\right) \;, & 
   y_d &= \left\vert \frac{\epsilon_1 \epsilon_2}{\epsilon_2 - \ci \, \tilde{\epsilon}_2}  \right\vert \;, \label{eq:LOquarkmixingA}\\ 
   \theta^{\text{CKM}}_{13}  &= \left\vert \frac{\epsilon_1}{\epsilon_3} \right\vert \;, &
   y_s &= \left\vert \epsilon_2 - \ci \, \tilde{\epsilon}_2 \right\vert \;, \label{eq:LOquarkmixingB} \\
   \theta^{\text{CKM}}_{23}  &= \left\vert \frac{\epsilon_2 + w \, \tilde{\epsilon}_2}{\epsilon_3} \right\vert \;, &
   y_b &= \left\vert \epsilon_3 \left(1 - \frac{1}{2} \hat{\zeta}_3^2 \right)\right\vert \;. \label{eq:LOquarkmixingC}
\end{align}
Note that we have to include the next-to-leading order term for $\theta^{\text{CKM}}_{12}$, which is comparatively large. 
Analogous terms for the other angles vanish due to the texture zeros in the third row of $Y_d$.
We can also derive an expression for the CKM phase $\delta_{\text{CKM}}$ in terms of the $\epsilon$'s
\begin{equation} \label{eq:modelCKMphase}
   \delta_{\text{CKM}} = -\arctan\frac{\tilde{\epsilon}_2}{\epsilon_2} \;,
\end{equation}
where we have dropped unphysical sign ambiguities in the $\epsilon$ parameters. Actually the only physical signs are the relative signs between $\epsilon_2$ and $\tilde{\epsilon}_2$ and the sign of $w$ leaving four possible sign combinations\footnote{However, requiring the correct CKM phase requires $w > 0$, which is already implied in Eq.\ \eqref{eq:modelCKMphase}. }. For simplicity we consider the solution where all $\epsilon$ parameters are positive except $\tilde{\epsilon}_2$.

\subsection{The Lepton Sector}

We discuss now the mixing angles and mass eigenvalues in the lepton sector. For the charged lepton sector we find at leading order
\begin{align}
   \theta^e_{12}  &= \left\vert \frac{c_2 \, \epsilon_2}{c_2 \, \epsilon_2 - \ci \, \tilde{c}_2 \, \tilde{\epsilon}_2} \right\vert \;, &
   y_e    &= \left\vert \frac{c_1 \, \epsilon_1 \, c_2 \, \epsilon_2}{c_2 \, \epsilon_2 - \ci \, \tilde{c}_2 \, \tilde{\epsilon}_2} \right\vert \;, \label{eq:LOleptonmixingA}\\ 
   \theta^e_{13}  &= 0 \;, &
   y_\mu  &= \left\vert c_2 \, \epsilon_2 - \ci \, \tilde{c}_2 \, \tilde{\epsilon}_2 \right\vert \;, \label{eq:LOleptonmixingB} \\
   \theta^e_{23}  &= 0 \;, &
   y_\tau &= \left\vert c_3 \, \epsilon_3 \left( 1 - \frac{1}{2} \hat{\zeta}_3^2 \right) \right\vert \label{eq:LOleptonmixingC} \;,
\end{align}
with the aforementioned ratios $c_1 = c_2 = c_3 = -3/2$ and $\tilde{c}_2 = 6$. Using also our formulae from the quark sector we can give relations for the GUT scale Yukawa coupling ratios
\begin{align}
   \frac{y_\tau}{y_b} &= \frac{3}{2} \;,\\
   \frac{y_\mu}{y_s}  &= 6 \sqrt{1 - \frac{15}{16} \cos^2 \delta_{\text{CKM}}} \approx 5.6 \;,\\
   \frac{y_e}{y_d}    &= \frac{9}{4} \frac{y_s}{y_\mu} \approx 0.4 \;,
\end{align}
where we especially want to note, that the ratio $y_\mu/y_s$ is not exactly equal to six as we naively imposed before, but is instead lowered by almost 10\% due to corrections from $\epsilon_2$. Nevertheless, all the ratios are still in good agreement with current experimental data \cite{Antusch:2009gu}.

For the neutrino mass matrix we find 
\begin{equation}
   m_\nu = \frac{\tilde{m}_2}{3} \begin{pmatrix} 1 & 1 & \left(1- \frac{1}{2} \hat{\zeta}^2_3 \right)\\ 1 & 1 & \left(1- \frac{1}{2} \hat{\zeta}^2_3 \right)\\ \left(1- \frac{1}{2} \hat{\zeta}^2_3 \right) & \left(1- \frac{1}{2} \hat{\zeta}^2_3 \right) & \left(1- \hat{\zeta}^2_3 \right) \end{pmatrix} + \frac{\tilde{m}_3}{2} \begin{pmatrix} 0 & 0 & 0\\ 0 & 1 & - \left(1- \frac{1}{2} \hat{\zeta}^2_3 \right)\\ 0 & -\left(1- \frac{1}{2} \hat{\zeta}^2_3 \right) & \left(1- \hat{\zeta}^2_3 \right) \end{pmatrix} \;,
\end{equation}
where $\tilde{m}_2 = \frac{3}{2} y_{\nu 1}^2 \frac{v_u^2}{M_{\nu^c,1}}$ and $\tilde{m}_3 = y_{\nu 2}^2 \frac{v_u^2}{M_{\nu^c,2}}$. The physical neutrino masses are then $m_2 = \tilde{m}_2 (1 - \frac{1}{3}\hat{\zeta}^2_3)$ and $m_3 = \tilde{m}_3 (1 - \frac{1}{2} \hat{\zeta}^2_3)$.
The lightest neutrino remains massless ($m_1 = 0$) in this minimal setup since we only introduced two right-handed neutrinos.

Using the approximate formulae of \cite{Antusch:2007ib,Antusch:2007vw,Antusch:2008yc}, we can obtain estimates for the mixing angles of the MNS mixing matrix as 
\begin{align}
 \sin \theta^{\text{MNS}}_{12} &= \frac{1}{\sqrt{3}} \left(1 + \frac{1}{6} \hat{\zeta}_3^2 \right) \;,  \label{eq:modelMNSangle12}\\
 \sin \theta^{\text{MNS}}_{23} &= \frac{1}{\sqrt{2}} \left(1 + \frac{1}{4} \hat{\zeta}_3^2 \right) \;, \label{eq:modelMNSangle23} \\
      \theta^{\text{MNS}}_{13} &= \frac{1}{\sqrt{2}} \left(1 + \frac{1}{4} \hat{\zeta}_3^2 \right) \frac{1}{\sqrt{1 + 16 \tan^2 \delta_{\text{CKM}}}}  \approx 0.06 \left(1 + \frac{1}{4} \hat{\zeta}_3^2 \right) \;, \label{eq:modelMNSangle13} \\
   \delta_{\text{MNS}} &= \pi + \arctan(4 \tan \delta_{\text{CKM}}) \approx \frac{3}{2} \pi \;,
\end{align} 
where corrections to the neutrino mixing parameters due to $\hat{\zeta}_3$ have been obtained using the approximate analytical formulae from \cite{Antusch:2008yc}, and sub-dominant corrections to $\theta_{12}^{\text{MNS}}$ and $\delta_{\text{MNS}}$ due to $\theta_{12}^e$ have been omitted. The model satisfies the generalized lepton mixing sum rule relations of \cite{Antusch:2007ib} (which compared to \cite{sumrule} also include third family  canonical normalization and RGE effects), which implies that the $\theta_{12}^e$-induced correction to $\theta_{12}^{\text{MNS}}$ is small due to small $\cos (\delta_{\text{MNS}})$.  
RGE corrections are subdominant as well (and have been omitted) since the neutrino mass spectrum is hierarchical and since the neutrino Yukawa couplings are much less than one \cite{Antusch:2005gp,Antusch:2003kp}. In particular, we see that our class of models always implies MNS mixings angles larger than the tribimaximal values. The prediction for $\theta_{13}^{\text{MNS}}$, for example, will be tested at future neutrino oscillation experiments \cite{Mezzetto:2009cr}.

We also note that neutrinoless double beta decay and direct determination of the neutrino mass in this class of models is beyond the reach of ongoing and planned experiments due to the small neutrino masses implied by the hierarchy and $m_1=0$. This could nevertheless be different in a variant of the model with a ``type II seesaw upgrade'' \cite{Antusch:2004xd,Antusch:2010es} lifting the restriction $m_1=0$.

\subsection{GUT Predictions and SUSY Threshold Corrections} \label{sec:scan}

Having the GUT scale structure of our model defined and using the simplified formulae for the SUSY threshold corrections as described in Sec.\ \ref{sec:SUSYThresholdCorrections}, we now derive constraints on the SUSY threshold corrections which we will then translate into bounds on the CMSSM parameter space.

\subsubsection{GUT Scale Yukawa Coupling Ratios}

We start our discussion of GUT scale Yukawa coupling ratios with the $b$-$\tau$ Yukawas:
\begin{equation}\label{eq:correctionOfbtauratio}
   \frac{3}{2} = \frac{y_\tau (M_{\text{GUT}}) }{y_b (M_{\text{GUT}})} \approx \frac{1 + (\epsilon_A + \epsilon_q)\tan\beta}{1+ \epsilon_l \tan\beta} \left.\frac{y_\tau (M_{\text{GUT}}) }{y_b (M_{\text{GUT}})}\right\vert_{\text{uncorrected}} \;,
\end{equation}
where we refer with ``uncorrected'' to the GUT scale values without SUSY threshold corrections included. For simplicity we apply them here at the GUT scale, which is a good first approximation, see \cite{Antusch:2008tf}. The value on the left side is the prediction of the model.

The uncorrected GUT scale value of this ratio can easily be obtained, e.g.\ using REAP \cite{Antusch:2005gp}. This yields the numerical value 
\begin{equation}\label{eq:uncorrectedbtauRatio}
   \left.\frac{y_\tau (M_{\text{GUT}}) }{y_b (M_{\text{GUT}}) }\right\vert_{\text{uncorrected}} = 1.29 \pm 0.05 - 0.03 \frac{\tan\beta}{30} \;,
\end{equation}
where the uncertainty is largely dominated by the bottom quark mass uncertainty and the dependence on $\tan\beta$ is a very good approximation
between $\tan\beta=10$ and $50$. Plugging this into Eq.\ \eqref{eq:correctionOfbtauratio} we find the relation:
\begin{equation}\label{eq:constraintAql}
   (\epsilon_A + \epsilon_q - \epsilon_l) \tan\beta \approx 0.16 \pm 0.05 + 0.03 \frac{\tan\beta}{30}\;.
\end{equation}
Similarly, we find for the second generation
\begin{equation}
   5.6 = \frac{y_\mu (M_{\text{GUT}}) }{y_s (M_{\text{GUT}})} \approx \frac{1 + \epsilon_q\tan\beta}{1+ \epsilon_l \tan\beta} \left.\frac{y_\mu (M_{\text{GUT}}) }{y_s (M_{\text{GUT}})}\right\vert_{\text{uncorrected}} \;.
\end{equation}
However, in this case the uncorrected GUT scale ratio has a much larger uncertainty due to the large strange quark mass uncertainty. The value and range obtained from REAP is then 
\begin{equation}
   \left.\frac{y_\mu (M_{\text{GUT}}) }{y_s (M_{\text{GUT}}) }\right\vert_{\text{uncorrected}} = 4.3^{+1.7}_{-1.0} \;, 
\end{equation}
which can be translated into the relation for the threshold corrections 
\begin{equation}\label{eq:constraintql}
   (\epsilon_q - \epsilon_l) \tan\beta \approx 0.32^{+0.38}_{-0.38} \;.
\end{equation}
Finally, for the first generation, we find
\begin{equation}
   0.4 = \frac{y_e (M_{\text{GUT}}) }{y_d (M_{\text{GUT}})} \approx \frac{1 + \epsilon_q\tan\beta}{1+ \epsilon_l \tan\beta} \left.\frac{y_e (M_{\text{GUT}}) }{y_d (M_{\text{GUT}})}\right\vert_{\text{uncorrected}} \;,
\end{equation}
with 
\begin{equation}
   \left.\frac{y_e (M_{\text{GUT}}) }{y_d (M_{\text{GUT}}) }\right\vert_{\text{uncorrected}} = 0.38^{+0.28}_{-0.11} \;, 
\end{equation}
which gives us the relation
\begin{equation}\label{eq:constraintql2}
   (\epsilon_q - \epsilon_l) \tan\beta \approx 0.05^{+0.44}_{-0.44} \;.
\end{equation}
This constraint is even weaker than the one for the second generation and therefore does not give us new insights. Both constraints are not particularly severe. 
Comparing the relations to the ranges given in Eqs.\ \eqref{eq:rangeLeptonThrCorr}--\eqref{eq:rangeBottomYukawaThrCorr} we expect that all constraints can be satisfied in a CMSSM-like setup.

From these ranges we can also derive a rough estimate for the optimal value of $\tan\beta$ needed to satisfy Eq.\ \eqref{eq:constraintAql}. Using the central values of the intervals of $\epsilon_q$, $\epsilon_A$ and $\epsilon_l$ this yields
\begin{equation}
   \tan\beta \approx 20 \;,
\end{equation}
from Eq.\ \eqref{eq:constraintAql}. In the more accurate analysis performed in \cite{Antusch:2009gu}, we see that a $b$-$\tau$ GUT scale Yukawa coupling ratio of $3/2$ (within the CMSSM) can only be obtained for $\tan\beta \approx 30$ or higher. For definiteness, we will fix $\tan\beta$ to the value
\begin{equation}
   \tan\beta = 30 \;,
\end{equation}
for our analysis.

\subsubsection{GUT Scale Mixing Angles}

For the mixing angles, we find equations similar to \eqref{eq:correctionOfbtauratio}, namely
\begin{equation}\label{eq:correctionOfi3angles}
 \theta_{i3}^{\text{CKM}}(M_{\text{GUT}}) = (1 + \epsilon_A \tan\beta) \left.\theta_{i3}^{\text{CKM}}(M_{\text{GUT}})\right\vert_{\text{uncorrected}} \;.
\end{equation}
Due to our matrix textures, we can always choose $w$ using Eq.\ \eqref{eq:LOquarkmixingC} to fit $\theta_{23}^{\text{CKM}}$. However, $\epsilon_1$ and $\epsilon_3$ are predetermined from the 1-2 sector and the third generation Yukawa coupling. To determine the model parameters, we can thus use the most constraining leading order equations, namely for $y_e$, $y_\mu$, $\theta_{12}^{\text{CKM}}$, $\theta_{23}^{\text{CKM}}$ and $y_\tau$ given in the previous two subsections. For this we use the numerical ``uncorrected'' GUT scale values and the approximation of applying SUSY threshold corrections at the GUT scale as before. The results are given in Tab.\ \ref{tab:numericalvaluesformodelparameters}. 

For the ``uncorrected'' value of $\theta_{13}^{\text{CKM}}$, i.e.\ the one obtained without including SUSY threshold corrections, we find 
\begin{equation}
   \left.\theta_{13}^{\text{CKM}}(M_{\text{GUT}})\right\vert_{\text{uncorrected}} = (3.24 \pm 0.15) \times 10^{-3} \;,
\end{equation}
and from Tab.\ \ref{tab:numericalvaluesformodelparameters} we get the model prediction at the GUT scale:
\begin{equation}
   \theta_{13}^{\text{CKM}}(M_{\text{GUT}}) = \frac{\epsilon_1}{\epsilon_3} = 3.62 \, \left( 1 - \frac{1}{2}\hat{\zeta}_3^2 \right) \times 10^{-3}  \;,
\end{equation}
where the dependence on $\epsilon_l \tan\beta$ has dropped out. Plugging this into Eq.\ \eqref{eq:correctionOfi3angles}, we find the relation
\begin{equation}\label{eq:constraintAzeta}
   \epsilon_A \tan\beta + \frac{1}{2}\hat{\zeta}_3^2 \approx 0.106 \pm 0.043 \;.
\end{equation}

\begin{table}
\begin{center}
\begin{tabular}{ccc}
\toprule
Quantity & Value\\
\midrule
$\epsilon_1$ in $10^{-4}$ & $5.56 \; q_l$ \\
$\epsilon_2$ in $10^{-4}$ & $6.36 \; q_l$ \\
$\tilde{\epsilon}_2$ in $10^{-3}$ & $-2.16 \; q_l$ \\
$\epsilon_3$ in $10^{-1}$ & $1.54 \; q_\tau$ \\
\midrule
$w$ & $ 2.66 \; q_\theta + 0.30$ \\
\bottomrule
\end{tabular}\end{center}
\caption{Numerical values for the GUT scale model parameters. The $q$ parameters are defined as $q_l = 1 + \epsilon_l \tan\beta$, $q_\tau = 1 + \epsilon_l \tan\beta - \frac{1}{2} \hat{\zeta}_3^2$ and $q_\theta = 1 + \frac{1}{2} \hat{\zeta}_3^2 + \epsilon_A \tan\beta$.
}\label{tab:numericalvaluesformodelparameters}
\end{table}

We can remove $\hat{\zeta}_3$ from this equation using the two neutrino mixing angles $\theta_{12}^{\text{MNS}}$ and $\theta_{23}^{\text{MNS}}$, i.e.\ Eqs.\ \eqref{eq:modelMNSangle12} and \eqref{eq:modelMNSangle23}. This gives 
\begin{equation}\label{eq:constraintAzetasimplified}
 \epsilon_A \tan\beta = 0.076^{+0.050}_{-0.053} \;.
\end{equation}
In addition, using the values in Tab.\ \ref{tab:numericalvaluesformodelparameters}, we derive
\begin{equation}
 \delta_{\text{CKM}} = 1.28 \;, 
\end{equation}
which is consistent with experiment within $1\sigma$. 

\subsection{Constraints on the CMSSM Parameter Space}

So far we have derived constraints on the SUSY threshold corrections from the demand to match our assumed GUT scale structure of the Yukawa matrices to the observed SM quantities.

In the next step we will use them to constrain the CMSSM parameter space that determines the SUSY threshold corrections. This implies that we neglect the small deviations from the CMSSM in the considered class of models. This is a good approximation for all flavon F-term parameters $C_{\phi_i}$ except $C_{\phi_3}$, which can enter the spectrum in non-negligible amount. However, it does so by lowering the right-handed sbottom and left-handed stau masses and changing the GUT scale $A_b$ and $A_\tau$ trilinear coupling parameters. The first two enter the threshold corrections only through loop functions, while the last two do not enter at all or only via running. Thus, barring extreme situations such as, for instance, large $C_{\phi_3}$, the threshold corrections should be quite insensitive to $C_{\phi_3}$.

To test which parts of the spectrum satisfy the equations \eqref{eq:constraintAql} and \eqref{eq:constraintAzetasimplified}, we implemented full one-loop RG evolution of the soft term parameters from the GUT scale to the SUSY scale. Instead of a fixed SUSY scale $M_{\text{SUSY}}$ we use the more common definition
\begin{equation}
  M_{\text{SUSY}} = \sqrt{m_{\tilde{t}_1} m_{\tilde{t}_2}} \;,
\end{equation}
where $m_{\tilde{t}_i}$ are the two stop masses. This scale is calculated via an iterative algorithm that repeats running between the GUT and the SUSY scale until $M_{\text{SUSY}}$ changes less than 1\%. At the SUSY scale the threshold corrections are included as described in Sec.\ \ref{sec:SUSYThresholdCorrections}.

Using this procedure we did a parameter scan for a CMSSM spectrum defined by the parameters $m_0$, $M_{1/2}$ and $A_0$. As ranges we took $m_0 \in [0; 4000]$~GeV, $M_{1/2} \in [100; 1500]$~GeV, $a_0 = A_0 / m_0 \in [1; 3]$. The motivation for the range of $A_0 / m_0$ is the fact that the contribution from gauginos to the running of trilinear terms always pulls them to smaller or even negative values of $A_t \propto \epsilon_A$.
Thus in order to satisfy Eq.\ \eqref{eq:constraintAzeta} and minimize deviation from tri-bimaximal neutrino mixing, we need to have a large positive value for $A_0$.
However, it is well known that $A_0$ cannot be set to arbitrarily
high value, to avoid charge and colour breaking (this requirement
typically gives $|A_0| \lesssim 3 m_0$) \cite{Frere:1983ag}. 
Furthermore, we chose $\mu > 0$ due to the phenomenological requirements of the anomalous magnetic moment of the muon, while $\tan\beta = 30$ was fixed as mentioned previously. 

\begin{figure}
\begin{center}
\includegraphics[scale=0.67]{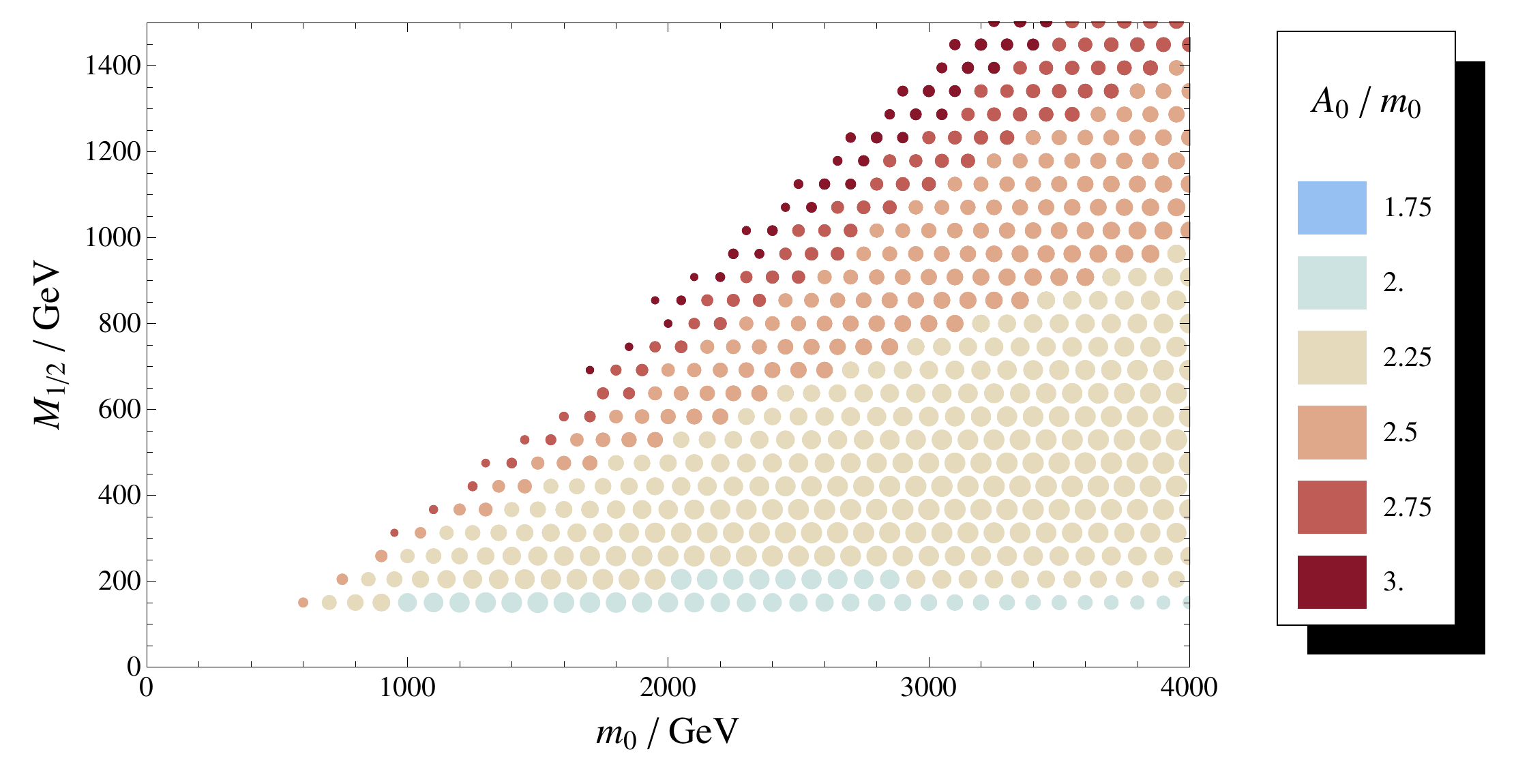}
\end{center}
\caption{This plot shows the sets of points satisfying the constraints in Eqs.\ \eqref{eq:constraintAql} and \eqref{eq:constraintAzetasimplified}. We only show the points with the least deviation for every given $m_0$ and $M_{1/2}$ in our scan. The size of the points is scaled with the deviation, such that larger points agree better and smaller points agree worse.
}\label{fig:spectrumscan}
\end{figure}

For our selection of points, we drop any that do not satisfy the LEP bounds on superpartner masses \cite{Amsler:2008zzb} or do not trigger successful EW symmetry  breaking, and also the ones where the lightest stau is the LSP, i.e.\ where it is lighter than the lightest neutralino. Out of the remaining set of points we simply select the ones satisfying \eqref{eq:constraintAql} and \eqref{eq:constraintAzetasimplified}. These are shown in Fig.\ \ref{fig:spectrumscan}. Note that  the recent results from ATLAS \cite{ATLAS} and CMS \cite{CMS} only give a constraint for the few points in the lower left corner of the allowed parameter space. These constraints are also valid for large $A_0/m_0$ and rather independent of $\tan \beta$, as discussed, e.g.\ in \cite{Allanach:2011ut}.

We would like to make a few remarks on the preferred parameter space:
\begin{itemize}
  \item We find that, generically, a value of $a_0 \gtrsim 2$ is preferred. 
  \item For lower values of $a_0$ the results depend only weakly on $m_0$ and we need a small $M_{1/2}$.
  \item Larger values of $M_{1/2}$ in general need larger values of $a_0$ and show a stronger correlation between $m_0$ and $M_{1/2}$.
  \item There is a lower bound on $m_0$ of about 600 GeV leading to a rather heavy SUSY spectrum. However, due to the large $a_0$ the third generation can still be relatively light. 
\end{itemize}

Note that, although at first glance our spectrum looks very similar to the ones necessary to achieve $b$-$\tau$ Yukawa coupling unification in the large $\tan \beta$ regime \cite{Blazek:2002ta}, this is indeed not the case. There, $a_0 = -2$ (note that the different sign is crucial here) and furthermore, although they also need $M_{1/2} \ll m_0$ as we do, the hierarchy there has to be even stronger (see for example \cite{Altmannshofer:2008vr, Baer:2009ff, Gogoladze:2009ug}).

Another interesting information is the range of achievable values for $\epsilon_A \tan \beta$. From the set of selected points we find that the largest possible value is $0.16$ and that only $2\%$ of the set has $\epsilon_A \tan \beta > 0.10$.

\subsection{Estimate of Flavour and CP Violating Effects}
\label{sec:estimate-fcnc}
In the following, we make use of the expressions presented in Sec.\ \ref{sec:flavor-effects} in order to provide 
an estimate of the main flavour and CP violation observables that will be subsequently studied numerically in Sec.\
\ref{sec:numerical}.

\subsubsection{LFV Processes}

Let us start considering $\mu\to e$ transitions. Within SUSY scenarios, the channel which currently provides the most stringent 
constraint to the $\mu$-$e$ sector is $\mu\to e\gamma$. Using the formulae given 
in Eqs.~(\ref{Eq:m2L-ll}, \ref{Eq:m2E-ll}, \ref{Eq:Adsckm}) with $\bar{m}_{\tilde \ell}\simeq m_0$, the leading contributions to the
$\mu$-$e$ mass insertions result:
\begin{align}
|(\delta^\ell_{\rm LL})_{21}| & ~\simeq~ \frac{1}{8 \pi^2} 
C_{2\tilde{2}}(C_{\tilde{\phi}_2} + a_0)\, c_{2}\epsilon_2 \,\tilde{c}_{2}\tilde{\epsilon}_2
\,\ln\frac{M_{\rm GUT}}{M_{\rm SUSY}} ~\simeq ~5\times 10^{-6}\,C_{2\tilde{2}}(C_{\tilde{\phi}_2} + a_0) \,, \label{eq:dLL12}\\
|(\delta^\ell_{\rm RR})_{21}|& ~\simeq~  \frac{1}{4 \pi^2} C_{\tilde{2}3}^2 w^2 \,c_{1}\epsilon_1 \,\tilde{c}_{2}\tilde{\epsilon}_2 
\,\ln\frac{M_{\rm GUT}}{M_{\rm SUSY}}~\simeq~ 10^{-4}\,C_{\tilde{2}3}^2\,,\label{eq:dRR12}\\
|(\delta^\ell_{\rm RL})_{21}| &~\simeq~  \frac{v_d}{m_0} C_{1\tilde{2}} \,\,c_{1}\epsilon_1
~\simeq~ 6\times 10^{-6} \left(\frac{1~{\rm TeV}}{m_0}\right) C_{1\tilde{2}}\,,\label{eq:dLR12}\\
|(\delta^\ell_{\rm LR})_{21}|& ~\simeq~  \frac{v_d}{m_0} C_{2\tilde{2}} \,\,c_{2}\epsilon_2
~\simeq~ 6\times 10^{-6} \left(\frac{1~{\rm TeV}}{m_0}\right) C_{2\tilde{2}}\,, \label{eq:dRL12}
\end{align}
where $a_0\equiv A_0/m_0$. 
The numerical estimates have been made by means of the values of the $\epsilon_i$ parameters as obtained from the semianalytical
analysis presented in the previous subsection (see Tab.~\ref{tab:numericalvaluesformodelparameters}) and the running of the 
$A^{\rm SCKM}_e$ entries has been neglected. 
As mentioned above, the only uncertainty in the full determination of the flavour violating effects relies on the flavon 
F-term coefficients $C_{\phi_{(i)}}$, which we assume to be $\mathcal{O}(1)$ \cite{Ross-Vives}. 
A comparison of the values in Eqs.~(\ref{eq:dLL12}-\ref{eq:dRL12}) with the bounds 
on the mass insertions in \cite{sleptonarium, paride} shows that the dominant contributions in the $\mu$-$e$ sector are provided
by the LR mixing insertions $(\delta^\ell_{\rm RL})_{21}$ and $(\delta^\ell_{\rm LR})_{21}$, i.e.\ directly by the off-diagonal entries
of $A^{\rm SCKM}_e$. This makes it particularly simple to estimate the resulting $\mu\to e\gamma$ rate.

On general grounds, the branching ratio of $\ell_{i}\to \ell_{j}\gamma$ can be written as:
\begin{equation}
\frac{{\rm BR}(\ell_{i}\to \ell_{j}\gamma)}{{\rm BR}(l_{i}\to
\ell_{j}\nu_i\bar{\nu_j})} =
\frac{48\pi^{3}\alpha}{G_{F}^{2}}(|A^{\rm L}_{ij}|^2+|A^{\rm R}_{ij}|^2)\,,
\label{eq:br}
\end{equation}
where the amplitudes $A^{\rm L}$, $A^{\rm R}$, in terms of LR mass insertions only, approximately read \cite{paride, Ciuchini:2007ha}:
\begin{equation}
A^{\rm L}_{ij} \simeq \frac{\alpha_1}{4 \pi} \frac{(\delta^\ell_{\rm RL})_{ij}}{\bar{m}^2_{\tilde \ell}} \frac{M_1}{m_{\ell_i}}
2 f(x_1)\,,~~ A^{\rm R}_{ij} = A^{\rm L}_{ij} ~({\rm L \leftrightarrow R})\,.
\end{equation}
Here $m_{\ell_i}$ is the {\it i\,}th generation lepton mass, $M_1$ is the Bino mass, $x_1 = M^2_1/\bar{m}^2_{\tilde \ell}$ and the loop
function $f(x)$ reads:
\begin{equation}
f(x) = \frac{1-5x^2+4x + 2x(x+2)\ln x}{4 (1-x)^4}\,.
\end{equation}
Inserting the values of $(\delta^\ell_{\rm RL})_{21}$ and $(\delta^\ell_{\rm LR})_{21}$ provided in Eqs.~(\ref{eq:dLR12}, \ref{eq:dRL12})
in the above formulae, we can finally get the following expression (which neglects subdominant contributions):
\begin{equation}
{\rm BR}(\mu\to e\gamma) \simeq 3.5\times 10^{-13} \left(C_{1\tilde{2}}^2 + C_{2\tilde{2}}^2\right) \left(\frac{1~{\rm TeV}}{m_0}\right)^4
\left(\frac{M_1}{100~{\rm GeV}}\right)^2\,,
\label{eq:br-est}
\end{equation}
where we assumed $M_1 \ll \bar{m}_{\tilde{\ell}}\simeq m_0$, a condition which is often verified within the model, 
as we will see in the next sections.
Remarkably, ${\rm BR}(\mu\to e\gamma)$ can be easily in the reach of the MEG experiment at PSI \cite{meg}, whose final sensitivity is
$\mathcal{O}(10^{-13})$. On the other hand, we see that just a conspiration of the $C_{\phi_{(i)}}$ coefficients can result in a
value of ${\rm BR}(\mu\to e\gamma)$ at the level of the present experimental limit ($1.2\times 10^{-11}$).\footnote{In the numerical 
analysis presented in Sec.\ \ref{sec:numerical}, we varied $|C_{\phi_{(i)}}|$ in the range $0.3\div3$, which corresponds to 
$|C_{1\tilde{2}}|,\, |C_{2\tilde{2}}| \leq 6$. Thus, according to Eq.~(\ref{eq:br-est}), we can get at most 
${\rm BR}(\mu\to e\gamma) \simeq 2.5\times 10^{-11}$.}

Another promising observable in the $\mu$-$e$ sector is represented by
$\mu\to e$ conversion in nuclei.
Within SUSY models, the latter process is typically dominated by the
same penguin diagram contributing to $\mu\to e \gamma$
(while a possible box diagram contribution turns out to be negligible)
with the photon attached to a proton line.
As a consequence, $\mu\to e$ conversion in nuclei just requires an
additional electromagnetic vertex.
We can then make use of Eq.~(\ref{eq:br-est}) to easily estimate the
conversion rate:
\begin{equation}
{\rm CR}(\mu\to e~{\rm in~N})\simeq \alpha_{\rm em}\times{\rm
BR}(\mu\to e\gamma)\,.
\end{equation}

Let's now consider the LFV in the $\tau$-$\mu$ sector, namely $\tau\to \mu\gamma$.
The approximate expressions for the $\tau$-$\mu$ mass insertions read:
\begin{align}
|(\delta^\ell_{\rm LL})_{32}| & ~\simeq~ (C_{\phi_3} \hat{\zeta}_{3})^2 (V^e_R)_{2 3} 
+\frac{1}{8 \pi^2} C_{\tilde{2}3}(C_{\tilde{\phi}_2} + a_0)\, w(\tilde{c}_{2}\tilde{\epsilon}_2)^2
\,\ln\frac{M_{\rm GUT}}{M_{\rm SUSY}} \nonumber \\
& ~\simeq~ 10^{-2} (C_{\phi_3} \hat{\zeta}_{3})^2 + 2.5\times 10^{-4}\,C_{\tilde{2}3}(C_{\tilde{\phi}_2} + a_0) \,, \label{eq:dLL23}\\
|(\delta^\ell_{\rm RR})_{32}|& ~\simeq~  \frac{1}{4 \pi^2} C_{\tilde{2}3}(C_{\phi_3} + a_0) w \,c_{3}\epsilon_3 \,\tilde{c}_{2}\tilde{\epsilon}_2 
\,\ln\frac{M_{\rm GUT}}{M_{\rm SUSY}}~\simeq~ 8\times 10^{-3}\,C_{\tilde{2}3}(C_{\phi_3} + a_0)\,,\label{eq:dRR23}\\
|(\delta^\ell_{\rm RL})_{32}| &~\simeq~  \frac{v_d}{m_0} C_{\tilde{2}3} \,\,w \tilde{c}_{2} \tilde{\epsilon}_2
~\simeq~ 2.5\times 10^{-4} \left(\frac{1~{\rm TeV}}{m_0}\right) C_{\tilde{2}3}\,,\label{eq:dLR23}\\
|(\delta^\ell_{\rm LR})_{32}|& ~\simeq~ \mathcal{O}\left(\frac{\tilde{c}_{2} \tilde{\epsilon}_2}{c_3\epsilon_3}\right)
\times  |(\delta^\ell_{\rm RL})_{32}| ~\simeq~ 0.05 |(\delta^\ell_{\rm RL})_{32}| \,. \label{eq:dRL23}
\end{align}
From these equations, we see that the main contribution to $|(\delta^\ell_{\rm LL})_{32}|\sim (C_{\phi_3} \hat{\zeta}_{3})^2 (V^e_R)_{2 3}$
does not arise from the running, but directly from the non-universality of the third generation LH sleptons, as discussed 
at the beginning of Sec.\ \ref{sec:flavor-effects}.
All the mass insertions can give a contribution to the $\tau\to \mu\gamma$ rate of a comparable size, so that the situation is more
involved than in the $\mu$-$e$ sector. The total rate of $\tau\to \mu\gamma$ should, however, be too small for detection at future experiments. For example, considering only $(\delta^\ell_{\rm RR})_{32}$, we can write the $A^{\rm R}_{32}$ amplitude like the following \cite{HNP}:
\begin{equation}
A^{\rm R}_{32} \simeq \frac{\alpha_1}{4 \pi} \frac{(\delta^\ell_{\rm RR})_{32}}{60} \frac{\tan\beta}{\bar{m}^2_{\tilde \ell}}\,.
\end{equation}
Using this expression with the mass insertion as in Eq.~(\ref{eq:dRR23}), Eq.~(\ref{eq:br}) then gives:
\begin{equation}
{\rm BR}(\tau\to \mu\gamma) \simeq 1.6\times 10^{-13} C_{\tilde{2}3}^2\left(C_{\phi_3} + a_0\right)^2 
\left(\frac{1~{\rm TeV}}{m_0}\right)^4 \left(\frac{\tan\beta}{30}\right)^2\,,
\label{eq:br32-est}
\end{equation}
which gives at most ${\rm BR}(\tau\to \mu\gamma) \sim \mathcal{O}(10^{-10})$. 
The other mass insertions give slightly smaller but comparable contributions. As we will see in Sec.\ \ref{sec:numerical}, a full
numerical evaluation shows that the different contributions can sum up to somewhat larger values of ${\rm BR}(\tau\to \mu\gamma)$
($\lesssim \mathcal{O}(10^{-9})$), but still barely within the sensitivity of future experiments.

In the $\tau$-$e$ sector, the rates are even smaller. For instance, we find ${\rm BR}(\tau\to e\gamma) \lesssim 10^{-11}$,
far beyond the reach of any foreseeable experiment.

\subsubsection{Electric Dipole Moments}
In a similar way we can estimate the size of the electric dipole moments as predicted by the model. 
From Eq.~(\ref{Eq:Adsckm}), we see that the A-term matrices, $A_d$ and $A_e$, acquire a CP violating phase in the 1-1 entry,
again as a consequence of the misalignment with respect to the corresponding Yukawas, induced by the flavon F-term coefficients. 
It is well known that complex A-terms can easily induce unacceptably large EDMs 
(this is often regarded as part of the ``SUSY CP problem''). It is therefore remarkable that, in our class of models, not only
the phases of the flavour changing entries but also the flavour conserving ones get suppressed by the expansion parameters $\epsilon_i$.
For instance, we see that the imaginary part of $(A^{\rm SCKM}_d)_{11}$ acquires an additional $\epsilon_2/\tilde{\epsilon}_2$ suppression
(i.e. of the order of the Cabibbo angle) with respect to the real part. This is even more effective in the leptonic sector, where
we have $(c_2 \epsilon_2)/(\tilde{c}_2 \tilde{\epsilon}_2)$, with $c_2 = -3/2$, $\tilde{c}_2 = 6$.
Complex A-terms clearly induce complex mass insertions through the RG running (see Eqs.\ \eqref{Eq:m2Q-ll}-\eqref{Eq:m2E-ll}). Therefore
flavour mixing induced sources of EDMs will be also present within our model.\footnote{For a recent discussion of flavour
violating contributions to EDMs, see for instance \cite{Hisano:2008hn}. For studies about flavoured CP violation within $SU(3)$ flavour models,
see \cite{su3edms}.}
Even though we are going to take into account all these contributions in the numerical analysis, let us here provide an estimate of the
electron EDM, as induced by a complex $(\delta^\ell_{\rm LR})_{11}$ only.

In the regime $M_1 \ll \bar{m}_{\tilde{\ell}}$, we can approximate the electron EDM like the following \cite{sleptonarium}:
\begin{equation}
\frac{d_e}{e} \simeq \frac{\alpha_1}{4 \pi} \frac{M_1}{\bar{m}^2_{\tilde \ell}} \Im\left[(\delta^\ell_{\rm LR})_{11}\right]\,,
\end{equation}
where the imaginary part of $(\delta^\ell_{\rm LR})_{11}$ can be obtained from Eq.~(\ref{Eq:Adsckm}):
\begin{equation}
\Im\left[(\delta^\ell_{\rm LR})_{11}\right]\simeq \frac{v_d}{\bar{m}_{\tilde \ell}} C_{2\tilde{2}} 
\frac{(c_2 \epsilon_2)^2 c_1 \epsilon_1}{(\tilde{c}_2 \tilde{\epsilon}_2)^2}\,. 
\end{equation}
Therefore, we get:
\begin{equation}
\frac{d_e}{e} \simeq  2.2\times 10^{-29} \left(\frac{M_1}{100~{\rm GeV}}\right)\left(\frac{1~{\rm TeV}}{m_0}\right)^3  C_{2\tilde{2}} 
 ~{\rm cm}\,,
\end{equation}
which gives at most  $d_e \sim \mathcal{O}(10^{-28})~e~{\rm cm}$, one order of magnitude below the present experimental limit. 
Therefore, the model predicts the electron EDM at a level that is at present consistent with non-observation, but still
several orders of magnitude above the SM prediction, such that it can be easily tested by the currently running experiments.

A similar situation is verified in the hadronic sector, where we have:
\begin{equation}
\Im\left[(\delta^d_{\rm LR})_{11}\right]\simeq \frac{v_d}{\bar{m}_{\tilde q}} C_{2\tilde{2}} 
\frac{\epsilon_2^2 \epsilon_1}{\tilde{\epsilon}_2^2} \simeq 3\times 10^{-7} \left(\frac{1~{\rm TeV}}{\bar{m}_{\tilde q}}\right)
C_{2\tilde{2}}\,, 
\end{equation}
which is about one order of magnitude below the present bound \cite{Isidori:2010kg}, given by neutron and atomic EDM searches.

\subsubsection{CP Violation in the Kaon Mixing}

Let us finally consider the bounds coming from SUSY contributions to the Kaon mixing, which are the most stringent ones
within the hadronic sector. As usual, the strongest constraint is given by the CP violation parameter $\epsilon_K$. In our model,
all the relevant phases can be derived from the expression for $A^{\rm SCKM}_d$ given in Eq.~(\ref{Eq:Adsckm}). As in the leptonic
sector, the largest contributions come directly from the off-diagonal entries of the A-term matrix.
The main contributions to $\epsilon_K$ are then given by:
\begin{align}
\Im\left[(\delta^d_{\rm LR})_{12}\right]& \simeq \frac{v_d}{\bar{m}_{\tilde q}} C_{2\tilde{2}} 
\frac{\epsilon_2 \epsilon_1}{\tilde{\epsilon}_2}\simeq 10^{-6} \left(\frac{1~{\rm TeV}}{\bar{m}_{\tilde q}}\right) C_{2\tilde{2}}\,, \\
\Im\left[(\delta^d_{\rm RL})_{12}\right]& \simeq \frac{v_d}{\bar{m}_{\tilde q}} C_{2\tilde{2}} 
\frac{\epsilon_2^2}{\tilde{\epsilon}_2}\simeq 10^{-6} \left(\frac{1~{\rm TeV}}{\bar{m}_{\tilde q}}\right) C_{2\tilde{2}}\,.
\end{align}
The resulting values are between one and two orders of magnitude below the present experimental bound, $\sim 10^{-4}$ \cite{Isidori:2010kg}.
Still, as we will see in Sec.\ \ref{sec:numerical}, the interplay among different contributions can provide sizeable deviations from
the SM prediction for $\epsilon_K$.

\section{Numerical Analysis}
\label{sec:num}

\subsection{Markov Chain Monte Carlo Analysis}
\label{sec:mcmc}

In the previous section we have discussed simple semi-analytical estimates, which showed that our class of models first of all seems to be compatible to current experimental data for fermion masses and mixings and second we have demonstrated how the CMSSM parameter space can be constrained.

In this section we will improve our analysis using a Markov Chain Monte Carlo (MCMC) analysis. From this we can derive posterior probability distributions for the model parameters based on the experimentally measured quantities and uncertainties. Furthermore, it allows us to incorporate the deviations from the CMSSM in the considered class of models and fit the resulting spectrum to the data. Finally we will discuss predictions for various observables derived from that.

\subsubsection{Procedure}

The MCMC analysis was done using a standard Metropolis-Hastings algorithm \cite{Amsler:2008zzb}. For the conditional probability density in the usual form $P(x|\theta)$ 
\begin{equation}
 P(x|\theta) \propto \exp\left(-\frac{\chi^2(\theta)}{2}\right),
\end{equation}
where $x$ represents measurements and $\theta$ model parameters, and where
with $\chi^2$ we denote the usual sum of deviations squared for the fitted parameters normalized with the experimental uncertainties. The quantities we use are the down-type quark and charged lepton Yukawa couplings, CKM mixing parameters and the 1-2 and 2-3 MNS mixing angles. Consequently, we varied only the $\epsilon_i$ parameters, $\hat{\zeta}_3$, $w$, $C_{\phi_3}$ and the SUSY breaking parameters $m_0$, $M_{1/2}$ and $a_0 = A_0/m_0$. The other parameters, i.e.\ the up-type quark Yukawa couplings and neutrino parameters were fixed to their respective GUT scale value obtained neglecting SUSY threshold corrections.

For the experimental values the quark and charged lepton masses at $m_t(m_t)$ were taken from \cite{Xing:2007fb} as input. The quark mixing angles were taken from the global fit parameters as given in \cite{Amsler:2008zzb}. For the parameters in the neutrino sector we used the values given in \cite{GonzalezGarcia:2010er}. In our analysis, we take the charged lepton Yukawa coupling uncertainties to be 1\% (although they are of course measured with higher precision). A higher accuracy would not be justified as we are using only one-loop RGEs, for instance.

These low energy observables are matched to the underlying model parameters to obtain the desired $\chi^2$ via the following procedure:
\begin{enumerate}
   \item Take the Yukawa matrices at the scale $M_{\text{GUT}}$ as shown in Eqs.\ \eqref{Eq:Yuku_can}-\eqref{Eq:Yuke_can}. 
   \item Evolve the Yukawa matrices from $M_{\text{GUT}}$ to $M_{\text{SUSY}}$ using the one-loop MSSM RGEs. Right-handed neutrinos are integrated out at their mass scale with the appropriate matching as given by the see-saw formula.
   \item Match MSSM and SM Yukawa couplings incorporating the SUSY threshold corrections as described in the previous section.
   \item Evolve the Yukawa matrices to the top mass scale $m_t(m_t)$ using SM RGEs.
\end{enumerate}

For the prior probability densities we used flat distributions for almost all varied parameters. Of these the following are notable: The massive parameters $m_0$ and $M_{1/2}$ were taken from the interval $[0; 4000]$ GeV and $[150;1500]$ GeV respectively. The prior for the trilinear coupling parameter $a_0=A_0/m_0$ was taken be flat on the interval $[0;3]$. Furthermore, we took $\hat{\zeta}_3^2$ instead of $\hat{\zeta}_3$ to be flatly distributed due to the observation that the relevant SM quantities depend linearly on the square of $\hat{\zeta}_3$. Thus a Gaussian distribution for these quantities also implies one for $\hat{\zeta}_3^2$ (neglecting boundary effects). The flavon F-term parameter $C_{\phi_3}$ was taken to be flatly distributed on the interval $[-3;3]$.

Based on this we calculated several Markov chains with the characteristics shown in Tab.~\ref{tab:semianalyticalMCMCstats}.

\begin{table}
\begin{center}
\begin{tabular}{lc}
\toprule
Quantity & Value \\
\midrule
Number of Chains & 5\\
Number of Distinct Points & 468238 \\
Number of Iterations & 7435125 \\
Minimal total $\chi^2$ with $n_{\text{dof}} = 2$ & 2.92 \\
$\text{PSRF} - 1$ & $4 \times 10^{-4}$ \\
\bottomrule
\end{tabular}\end{center}
\caption{Statistical characteristics of the Markov chains. 
PSRF is the maximal potential scale reduction factor \cite{gelman1992inference} over all parameters.
}\label{tab:semianalyticalMCMCstats}
\end{table}

\subsubsection{Results}

The standard interpretation of the points found by the Markov chains is that they are a sample from the posterior probability distribution $P(\theta|x)$. We can thus give probabilities for the model parameters and therefore $68\%$ intervals and multidimensional boxes, where the parameters lie with that probability. The values obtained from our analysis are shown in Tab.~\ref{tab:fullMCMCvalues}. 
We did not show values for the $C_{\phi_3}$ and massive parameters $m_0$ and $M_{1/2}$ as they all have a rather flat distribution, i.e.\ excess kurtosis\footnote{For comparison, we note that a completely flat distribution would correspond to an excess kurtosis of $-1.2$.} $\le -1$, over the whole range allowed by the prior. However, the Markov chains prefer heavy masses and disfavour large positive $C_{\phi_3}$.

Based on this sample, we can also derive a probability distribution for each fitted SM quantity, interpreting them as functions and thus new random variables dependent on the model parameters. In the best case scenario they should be centered exactly on the experimental input value with a standard deviation matching the experimental uncertainty (or the one put into the analysis) and naturally be normally distributed. Deviations from this can naturally arise through model inherent correlations, prior effects and lacking convergence of the MCMC data distribution. 
The quantities obtained from our analysis are shown in Tab.~\ref{tab:MCMCresults}.

\begin{table}
\begin{center}
\begin{tabular}{ccc}
\toprule
Parameter & Value with $68\%$ Range\\
\midrule
$\epsilon_1$ in $10^{-4}$ & $5.88 \pm 0.11$\\
$\epsilon_2$ in $10^{-4}$ & $6.64 \pm 0.13$\\
$\tilde{\epsilon}_2$ in $10^{-3}$ & $2.27 \pm 0.04$\\
$\epsilon_3$ in $10^{-1}$ & $1.673 \pm 0.073$\\
\midrule
$\frac{1}{2}\hat{\zeta}_3^2$ & $0.035 \pm 0.033$\\
$w$ & $3.14 \pm 0.12$ \\
\midrule
$a_0$ & $2.13 \pm 0.51$\\
\bottomrule   
\end{tabular}
\end{center} 
\caption{Distribution characteristics of model parameters as found by the full numerical MCMC analysis.}\label{tab:fullMCMCvalues}
\end{table}

\begin{table}
\begin{center}
\begin{tabular}{cccc}
\toprule
Parameter & Experiment & MCMC $68$\% Range  \\
\midrule
$\theta_{12}^{\text{CKM}}$ & $0.2277 \pm 0.0010$ & $0.2276 \pm 0.0010$ \\
$\theta_{23}^{\text{CKM}}$ & $0.0414 \pm 0.0012$ & $0.0414 \pm 0.0012$ \\
$\theta_{13}^{\text{CKM}}$ & $(3.60 \pm 0.17) \times 10^{-3}$ & $(3.76 \pm 0.12) \times 10^{-3}$ \\
$\delta_{\text{CKM}}$ & $1.202^{+0.078}_{-0.043}$ & $1.286 \pm 0.003$ \\
\midrule
$y_d$ in $10^{-5}$ & $1.59 \pm 0.68$ & $1.62 \pm 0.53$ \\
$y_s$ in $10^{-4}$ & $2.99 \pm 0.86$ & $2.63 \pm 0.86$ \\
$y_b$ in $10^{-2}$ & $1.579 \pm 0.051$ & $1.564 \pm 0.047$ \\
\midrule
$y_e$ in $10^{-6}$ & $2.78$ & $2.78 \pm 1\%$ \\
$y_\mu$ in $10^{-4}$ & $5.88 $ & $5.84 \pm 1\%$ \\
$y_\tau$ in $10^{-2}$ & $1.00 $ & $1.00 \pm 1\%$ \\
\midrule
$\theta_{12}^{\text{MNS}}$ & $0.602 \pm 0.017$ & $0.625 \pm 0.008$ \\
$\theta_{23}^{\text{MNS}}$ & $0.749 \pm 0.071$ & $0.825 \pm 0.018$ \\
$\theta_{13}^{\text{MNS}}$ & -- & $0.053 \pm 0.001$ \\
$\delta_{\text{MNS}}$ & -- & $4.81 \pm 0.05$ \\
\bottomrule
\end{tabular}\end{center}
\caption{SM quantities obtained from the MCMC data. 
The minimal $\chi^2$ (including down-type quarks and charged lepton Yukawa couplings, CKM parameters, $\theta_{12}^{\text{MNS}}$ and $\theta_{23}^{\text{MNS}}$) is $2.92$ with two degrees of freedom corresponding to a p-value of $23$\%. 
}\label{tab:MCMCresults}
\end{table}

Looking at those numbers, we see that the considered class of models can in general describe the measured SM quantities quite good. The largest deviations are a very pronounced value for $\delta_{\text{CKM}}$ at the $1\sigma$ interval boundary, which can be attributed to the highly constrained 1-2 sector of the Yukawa matrices $Y_d$ and $Y_e$. Then there is a shift of the value of $\theta_{13}^{\text{CKM}}$ also by about $1\sigma$ and the neutrino mixing angles by $1\sigma$, each however with large standard deviations. This is expected from the range of possible values for $\epsilon_A$ together with the form of the constraint given by Eq.\ \eqref{eq:constraintAzeta}, which forces us to choose between a deviation in $\theta_{13}^{\text{CKM}}$ and a deviation in the neutrino mixing angles. Analogously, this also explains the lower value for $w$ compared to the value given in Tab.\ \ref{tab:numericalvaluesformodelparameters}, since it does not have to counter-act such a large $\epsilon_A \tan \beta + \frac{1}{2}\hat{\zeta_3}^2$ to arrive at an acceptable $\theta_{23}^{\text{CKM}}$.

Another possible interpretation of the Markov chain data is that it just constitutes a sophisticated scan over the parameters that happens to be more fine-grained in the interesting regions of low $\chi^2$. We can thus select from the chains those points that yield a $\chi^2$ lower than some threshold value and demote the Bayesian probability interpretation to the nice side-effect of enhanced resolution in parts of the parameter space\footnote{Strictly speaking this is not necessarily true because the MCMC algorithm can stay at the same point for more than one iteration resulting in a possibly crucial difference between point and probability density distribution.}.

\subsubsection[Constraints on $m_0$, $M_{1/2}$ and $A_0$]{Constraints on $\boldsymbol{m_0}$, $\boldsymbol{M_{1/2}}$ and $\boldsymbol{A_0}$}

\begin{figure}
\begin{center}
\includegraphics[scale=0.67]{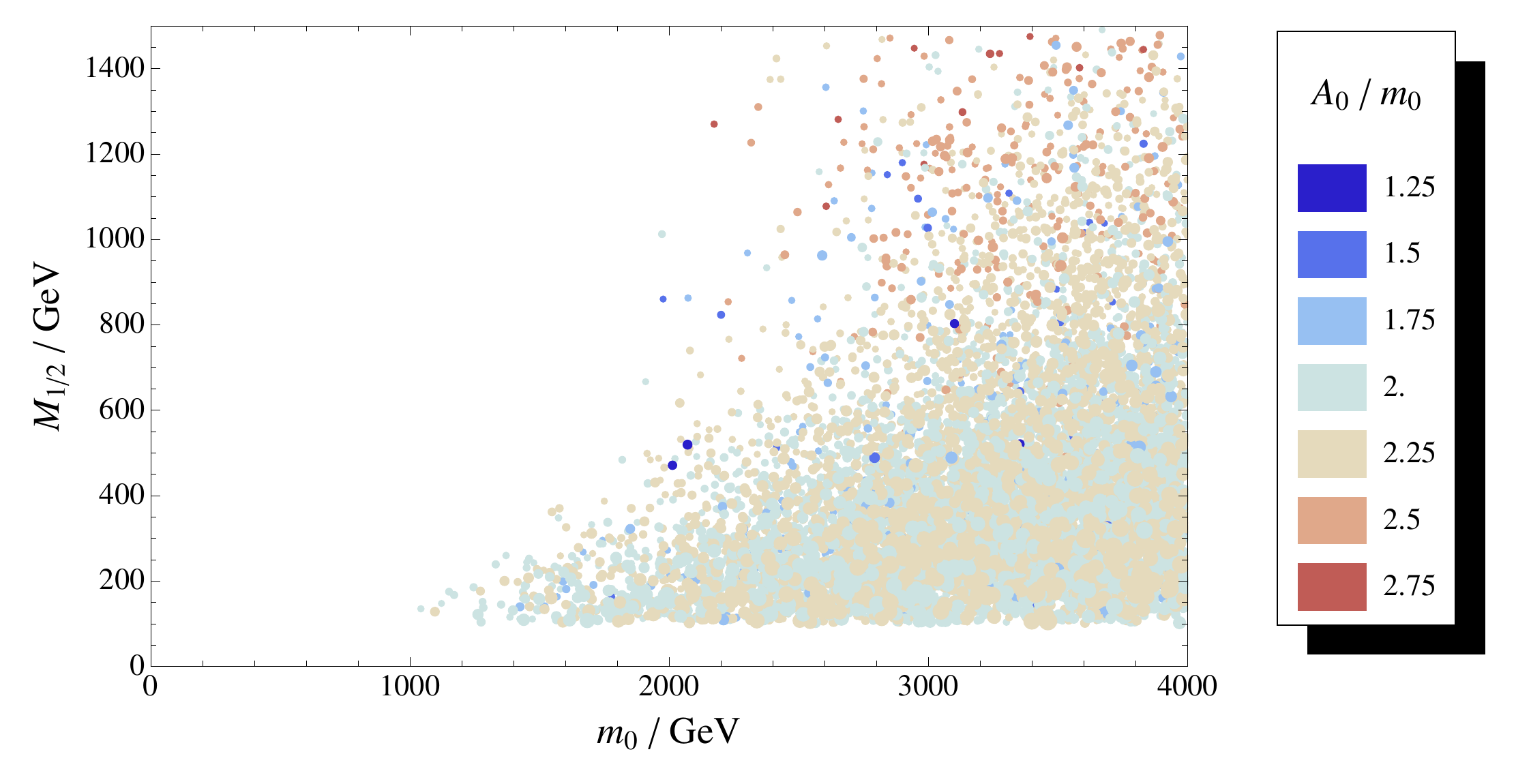}
\end{center}
\caption{Points found by the Markov chains with $\chi^2 < 5.99$ (p-values larger than $0.05$). Points are ordered with respect to their $\chi^2$ such that the points which are fitting the experimental values better are drawn on top. In addition, point size scales with $\chi^2$, making points agreeing better with experimental data larger.
}\label{fig:fullMCMCspectrum}
\end{figure}

Based on the full MCMC analysis, we can also compute the constraints on $m_0, M_{1/2}$ and $A_0$, in analogy to the plot shown in Fig.\ \ref{fig:spectrumscan} where we used the analytical approximations. The set of points with $\chi^2 < 5.99$ is shown in Fig.\ \ref{fig:fullMCMCspectrum}. With the two residual degrees of freedom of our fit this corresponds to p-values larger than $0.05$.

We can see that both spectrum plots look fairly similar. One difference,
however, is that in Fig.\ \ref{fig:fullMCMCspectrum} we find light orange as well as light blue points in the central region of the points, whereas in the plot in Fig.\ \ref{fig:spectrumscan} there are mainly light orange points. This is explained by the fact that the colour of the points in Fig.\ \ref{fig:spectrumscan} indicates the value of $A_0/m_0$ which provides the best fit - and $A_0/m_0 = 2.25$ turned out to be slightly better in this respect than $A_0/m_0 = 2$ - whereas in Fig.\ \ref{fig:fullMCMCspectrum} all points of the MCMC sample with $\chi^2 < 5.99$ are shown.
Another difference can be seen in the lower
bound on $m_0$. It moves from 600 GeV to 1 TeV due to the fact that
the points on the edge of the allowed parameter space of Fig.\ \ref{fig:spectrumscan}
need larger values of $1/2 \hat{\zeta}_3^2$ which are
allowed by the approach used for the scan but disfavoured by the
experimental values of the neutrino angles. This leads to an exclusion
of the points with lower $m_0$
by the requirement $\chi^2 < 5.99$.

Thus we can conclude that our approach in the simplified parameter scan
is useful to get a first impression of the allowed parameter space.
However, we stress that a full numerical MCMC analysis is still
necessary to obtain accurate constraints, e.g.\ for $m_0$. Furthermore
it carries additional useful probabilistic information, e.g.\ in the apparent
point density. 

%

\subsection{Phenomenological Consequences} 
\label{sec:numerical}

In this section, we present the results of a numerical analysis performed 
by scanning the parameter space favoured by the MCMC fit presented above.
The SUSY parameters $m_0$, $M_{1/2}$, $A_0$, as well as $\epsilon_{(i)}$, 
$w$, $\hat{\zeta}_3$ and $C_{\phi_3}$ were taken from the MCMC fit discussed 
in the previous section.
We remind that for the fit we used $\tan\beta =30$.
The coefficients relating flavon F-terms and flavon vevs were then randomly varied 
in the following ranges:
\begin{align}
0.3 \leq |C_{\phi_1}| \leq 3\,,~~0.3 \leq |C_{\phi_2}| \leq 3\,, ~~
0.3 \leq |C_{\tilde{\phi}_2}| \leq 3\,.
\end{align}

Following the discussion of Sec.\ \ref{sec:normalisedYukawas}, the parameters 
$\zeta_{(i)}^{24}$ have been assumed to be $\mathcal{O}(1)$, with the exception of  
$\zeta_{\tilde 2}^{24}$ which has been set to $0.1$. 

In order to compute the SUSY spectrum,
we numerically solve the full 1-loop RGEs of the MSSM (with 2-loop RGEs for gauge couplings
and gaugino masses) from $M_{\rm GUT}$ down to the SUSY scale 
$M_{\rm SUSY} \equiv \sqrt{m_{\tilde{t}_1} m_{\tilde{t}_2}}$.\footnote{We notice that a full 2-loop computation might be necessary for a
correct evaluation of the spectrum, especially for regions of the
parameter space where EWSB requires precise tuning. However, we
checked that for most of the parameter space 1-loop RGEs with the
inclusion of 2-loop contributions to the $\beta$ functions of the gauge couplings and gaugino
masses is accurate enough for the estimates we present here.}
For each point of the parameter space, we impose 
the following requirements: (i) successful EWSB and absence of tachyonic particles;
(ii) limits on SUSY masses from direct searches; (iii) neutral LSP.
Then, we compute the leptonic processes by means of a full calculation
in the mass eigenstate basis \cite{lepton-processes}, 
the hadronic processes by means of the mass-insertion 
approximation formulae in \cite{MIs,Altmannshofer:2009ne},
and the ${\rm BR}(B\to X_s \gamma)$ using \texttt{SusyBSG} \cite{susybsg}.
We then require that the resulting ${\rm BR}(B\to X_s \gamma)$ does not deviate from 
the experimental value \cite{HFAG} in more than 3$\sigma$ 
($2.75\times 10^{-4}\leq{\rm BR}(B\to X_s \gamma)\leq 4.25\times 10^{-4}$)
and that ${\rm BR}(B_s\to \mu^+ \mu^-) \leq 5.2 \times 10^{-8}$, 
$0.77\leq |\epsilon_K/\epsilon_K^{\rm exp}|\leq 1.23$ 
\cite{Altmannshofer:2009ne,Buras:2010pm}. 
The LEP limit $m_h \geq 114$ GeV does not impose additional 
constraints to the parameter space.

\subsubsection{Spectrum}

\begin{figure}[t]
\centering
\includegraphics[width=0.55\textwidth]{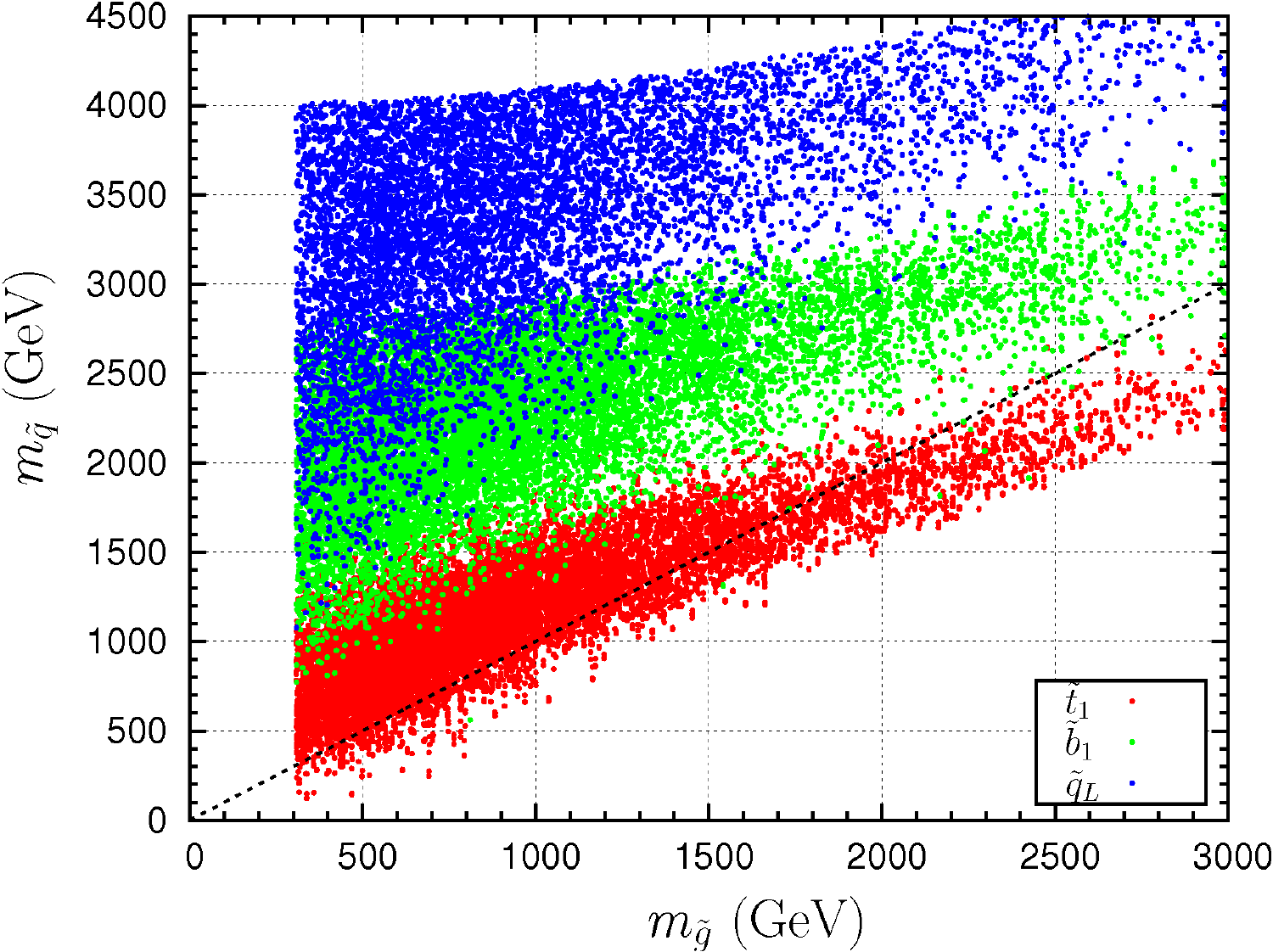}
\caption{Squark (blue: $\tilde{q}_L$; green: $\tilde{b}_1$; red: $\tilde{t}_1$) versus gluino masses from the MCMC analysis. 
See the text for details.\label{Fig:mstop-mgluino}}
\end{figure}

\begin{table}[t]
\begin{center}
\begin{tabular}{cccc}
\toprule
Particle & Mass Range (GeV) & Particle & Mass Range (GeV) \\
\midrule
$\tilde g$ & $308\div 3177$ & ${\tilde \chi}^0_1$ & $55\div 685$  \\ 
${\tilde t}_1$ & $104\div 2908$ & ${\tilde \chi}^+_1$ & $107\div 1267$  \\ 
${\tilde b}_1$ & $552\div 3738$ & ${\tilde \tau}_1$ &  $383\div 3680$     \\
${\tilde q}$ & $1077\div 4807$ & ${\tilde \ell}$ &  $1048\div 4091$     \\  
$h$ & $115\div 129$ & $A$ &  $261 \div 4480$     \\  
\bottomrule
\end{tabular}\end{center}
\caption{SUSY mass ranges used in the analysis. The upper values just reflect the choice of scanning up to
$m_0 = 4$ TeV, $M_{1/2} = 1.5$ TeV. On the contrary, the lower values result from the successful 
fit of fermion masses and mixings and/or the applied constraints (see the text for details).
\label{tab:spectrum}}
\end{table}

The MCMC sample we used for this analysis corresponds to the 
ranges for the SUSY particle masses reported in Tab.~\ref{tab:spectrum}.
The lower values shown in the table give an indication of the SUSY spectrum which can better account for the 
fermion masses and mixing: in the case of sfermions they correspond to the fact that for a given 
value of $M_{1/2}$ a minimum value for $m_0$ is selected by the fit (as depicted in Fig.~\ref{fig:fullMCMCspectrum}).
On the contrary, the upper values are just related to our choice of scanning the parameter space up to $m_0 = 4$ TeV,
$M_{1/2} = 1.5$ TeV.

As we can see a rather heavy spectrum is selected, e.g.\ the heavier squarks are always above the TeV scale. 
However, gluinos and lightest stop and sbottom (${\tilde t}_1$, ${\tilde b}_1$) can be much lighter. 
In particular, $m_{{\tilde t}_1}$ is driven to light values by the large value $A_0 \gtrsim 2 m_0$ required by the fit. 
In Fig.~\ref{Fig:mstop-mgluino}, we show the heavy squarks mass (blue) and the third generation squark masses 
$m_{{\tilde b}_1}$ (green) and $m_{{\tilde t}_1}$ (red) versus the gluino mass $m_{\tilde g}$ 
for the scanning of the parameter space described above. 
We see that the squarks turn out to be always heavier (even much heavier) than gluinos in 
the SUSY parameter space region favoured by our model, with the possible exception of ${\tilde t}_1$, 
especially for large values of $M_{1/2}$ (i.e. of $m_{\tilde{g}}$). 

Even if gluinos might be light enough for an early SUSY discovery,\footnote{With 1 ${\rm fb}^{-1}$ of collected data, 
the discovery potential of the 7 TeV LHC run, 
might reach $m_{\tilde g} \sim 620$ GeV, in the case, like in our scenario, of $m_{\tilde g} \ll m_{\tilde q}$ \cite{Baer2010}.}
it is clear that such an eventuality would 
select a corner of the parameter space (the bottom-left one in Fig.~\ref{fig:fullMCMCspectrum}). 
It is remarkable that the spectrum itself gives the possibility of testing our class of flavour models and distinguishing them
from other scenarios. In this sense, the heavy spectrum, the squarks heavier than gluinos 
and the relatively light third generation are all generic features of our scenario, which the LHC could test after enough years of data taking.

Sleptons are usually heavy with the possible exception of the lightest stau. 
In fact, we recall that most of the parameter space points lie in the region where 
$m_0 \gg M_{1/2}$
and thus $m_{\tilde \ell}\simeq m_{\tilde q}$, since relatively light gluinos do not drive the squark masses 
to much larger values than sleptons in the RG running. 
In particular, the slepton masses turn out to be always larger than the mass of the $\tilde W$ (and thus of the 
lightest chargino and the second lightest neutralino). This means that they cannot be efficiently
produced in cascade decays of squarks to neutralinos. On the other hand, 
the direct Drell-Yan production cross-section for $m_{\tilde \ell}\gtrsim 200 \div 300$ GeV might be  
too small to have a slepton signal above the SM background \cite{Drell-Yan}.  As a consequence, sleptons seem to be hardly
detectable at the LHC, within our scenario.
However, the lightest stau represents an exception also in this case, since in some regions of the parameter space
$m_{{\tilde \tau}_1} < m_{{\tilde \chi}^0_2}$ and the decay ${\tilde \chi}^0_2 \to {\tilde \tau}_1 \tau \to {\tilde \chi}^0_1
\tau \tau $ is open and potentially observable at the LHC.

In our scenario, not only a specific region of the CMSSM parameter space is selected, but also effects from $C_{\phi_3}$
may result in non-negligible differences compared to a CMSSM spectrum. Observation of such a $C_{\phi_3}$ signature
would of course be interesting, however, since only the sbottom and the stau masses are affected 
(cf. Eqs.~(\ref{Eq:m2d}, \ref{Eq:m2l}) and Eqs.~(\ref{Eq:Ad}, \ref{Eq:Ae})), this seems to be experimentally challenging. 

The LSP is mostly Bino over all the parameter space. This is because the large $A_t$ required by the fit 
drives the running of $m^2_{H_u}$ to large negative values so that the Higgsino mass parameter 
$\mu$ results always quite large (in particular $\mu\gg M_1$), once the EWSB condition, $\mu^2 \simeq - m^2_{H_u}$, is imposed.
In fact, we find $\mu \gtrsim 650$ GeV over all the parameter space and consequently a quite small Higgsino component 
of the lightest neutralino ($\lesssim 0.1$).
As a consequence, the LSP annihilation cross-section is not enhanced by a sizeable Higgsino component, and the so-called focus-point
region for neutralino dark matter \cite{focuspoint} is not viable in this scenario (for $\tan\beta=30$).
On the other hand, there are some points of the parameter space (giving $m_{{\tilde \tau}_1} \simeq m_{{\tilde \chi}^0_1}$), 
where an efficient stau-neutralino coannihilation  \cite{staucoann} can reduce the neutralino relic density within the WMAP bound. 
This is because, as noticed above, the lightest stau $\tilde{\tau}_1$ can be driven light for sizeable values of $A_0$ and $C_{{\phi}_3}$, 
for which $A_{\tau}$ gets increased (cf. Eq.~(\ref{Eq:Ae})).
Moreover, the non-universal contribution to $m^2_{H_d}$, shown in Eq.~(\ref{Eq:m2Hd}), can be used to lower the mass
of the CP-odd Higgs, $A$, triggering the so-called A-funnel region (corresponding to $m_A\simeq 2 m_{{\tilde \chi}^0_1}$), where the LSP annihilates efficiently through an s-channel $A$ exchange \cite{funnel}. 
In our model, this opens up the possibility to access the funnel region even for moderate values of $\tan\beta$.

\subsubsection{Flavour observables}

\begin{figure}
\centering
\includegraphics[width=0.55\textwidth]{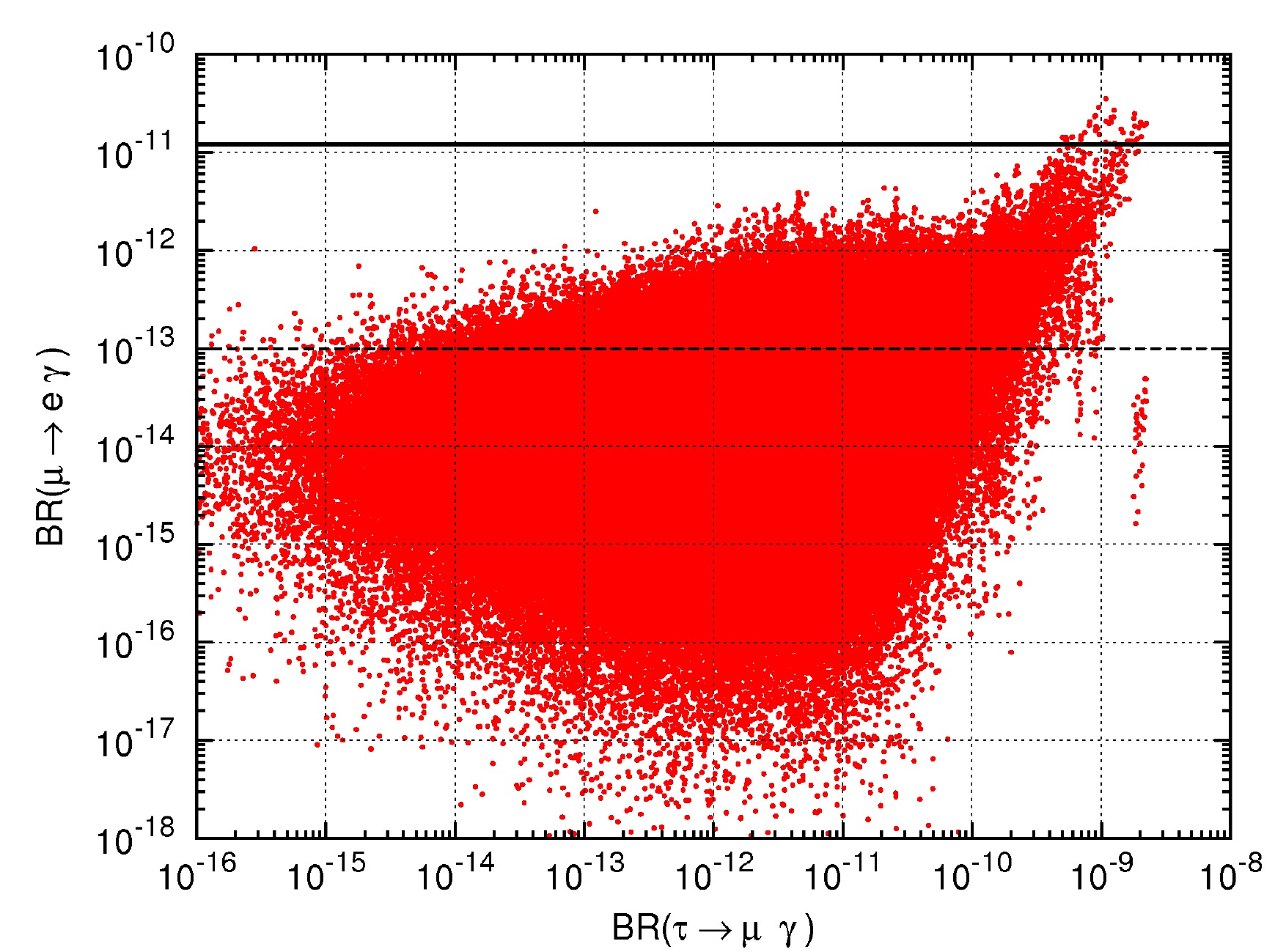}
\caption{BR$(\mu\to e\gamma)$ versus BR$(\tau\to \mu\gamma)$. The thick black line represents the current experimental 
limit BR$(\mu \to e\gamma) < 1.2\times10^{-11}$ \cite{MEGA}, the dashed line the expected sensitivity of the MEG experiment.\label{Fig:meg-tmg}}
\end{figure}

Let us now consider the predicted rates for the LFV observables $\mu\to e\gamma$ and
$\tau\to \mu\gamma$. In Fig.~\ref{Fig:meg-tmg}, we plot BR$(\mu\to e\gamma)$ versus BR$(\tau\to \mu\gamma)$, 
for the variation of the parameters (in particular the $C_{\phi_{(i)}}$) described above. 
We see that our class of models is at present not much constrained by the experimental limit on BR$(\mu\to e\gamma)$, 
which is shown in the figure as a black thick line. Only a tiny portion of the parameter space, 
corresponding to a rather light SUSY spectrum and a conspiration of the $|C_{\phi_{(i)}}|$ parameters, is already excluded.
This result is consistent with the semi-analytical estimate provided in Eq.~(\ref{eq:br-est}).
The final sensitivity reach on BR$(\mu\to e\gamma)$ of the MEG experiment \cite{meg} is represented by a black dashed line.
We see that MEG is able to test indirectly a large part of the parameter space. 
As expected by the discussion in Sec.\ \ref{sec:estimate-fcnc}, BR$(\tau\to \mu\gamma)$ cannot be as large as 
$\mathcal{O}(10^{-8})$, thus resulting beyond the reach of the SuperB factory at KEK \cite{KEK}. 
A positive signal for $\tau\to \mu\gamma$ would then disfavour our class of models.
There is a corner of the parameter space which can be in the reach of a Super Flavour factory at the level of 
BR$(\tau\to \mu\gamma)\simeq 10^{-9}$ \cite{superF}. Interestingly, these few points basically correspond to 
BR$(\mu\to e\gamma)>\mathcal{O}(10^{-13})$ and will be then tested by MEG, so that a negative result would preclude
the possibility of observing $\tau\to \mu\gamma$ at a Super Flavour factory as well.

The fact that still a wide portion of the points in Fig.~\ref{Fig:meg-tmg} would escape the searches at MEG is a consequence 
of our choice of scanning the parameter space up to very heavy slepton masses ($\simeq 4$ TeV). However, it is worth notice
that even such a heavy spectrum can provide rates of $\mu\to e\gamma$ in the reach of MEG. 
Moreover, there are very good prospects to test most of the points of Fig.~\ref{Fig:meg-tmg} in the future experiments
looking for $\mu\to e$ transition in nuclei.
We remind that, within SUSY models, there is typically a striking correlation between $\mu\to e\gamma$ and the $\mu\to e$ conversion
in nuclei, namely the $\mu\to e$ conversion rate is well approximated by 
${\rm CR}(\mu\to e~{\rm in~N})\simeq \alpha_{\rm em}\times{\rm BR}(\mu\to e\gamma)$. This means that
our prediction for BR$(\mu\to e\gamma)$ can be easily translated in a prediction for 
${\rm CR}(\mu\to e\,{\rm in\, N})$. The proposed $\mu-e$ conversion experiments at Fermilab~\cite{Carey:2008zz}
and at J-PARK~\cite{prism} aim at the respective sensitivities of $10^{-16}$
and $10^{-18}$ on ${\rm CR}(\mu\to e~{\rm in~ Ti})$. Therefore, such experiments would access the parameter 
space displayed in Fig.~\ref{Fig:meg-tmg} almost completely, thus testing the model well beyond the MEG reach.

As for LFV processes at colliders, such as $\tilde{\chi}_2^0 \to \tilde{\chi}_1^0 \mu e$, we notice that they are very much suppressed
in our scenario. This is due to the spectrum selected by the fermion masses and mixings fit, for which sleptons are heavy, 
and the second neutralino $\tilde{\chi}_2^0$ mass is always smaller than the slepton ones, apart from some points with a 
light $\tilde{\tau}_1$. Therefore, the LFV neutralino decay is a 3-body decay
with off-shell intermediate sleptons, which give a GIM-like
suppression of the process.

\begin{figure}
\centering
\includegraphics[width=0.47\textwidth]{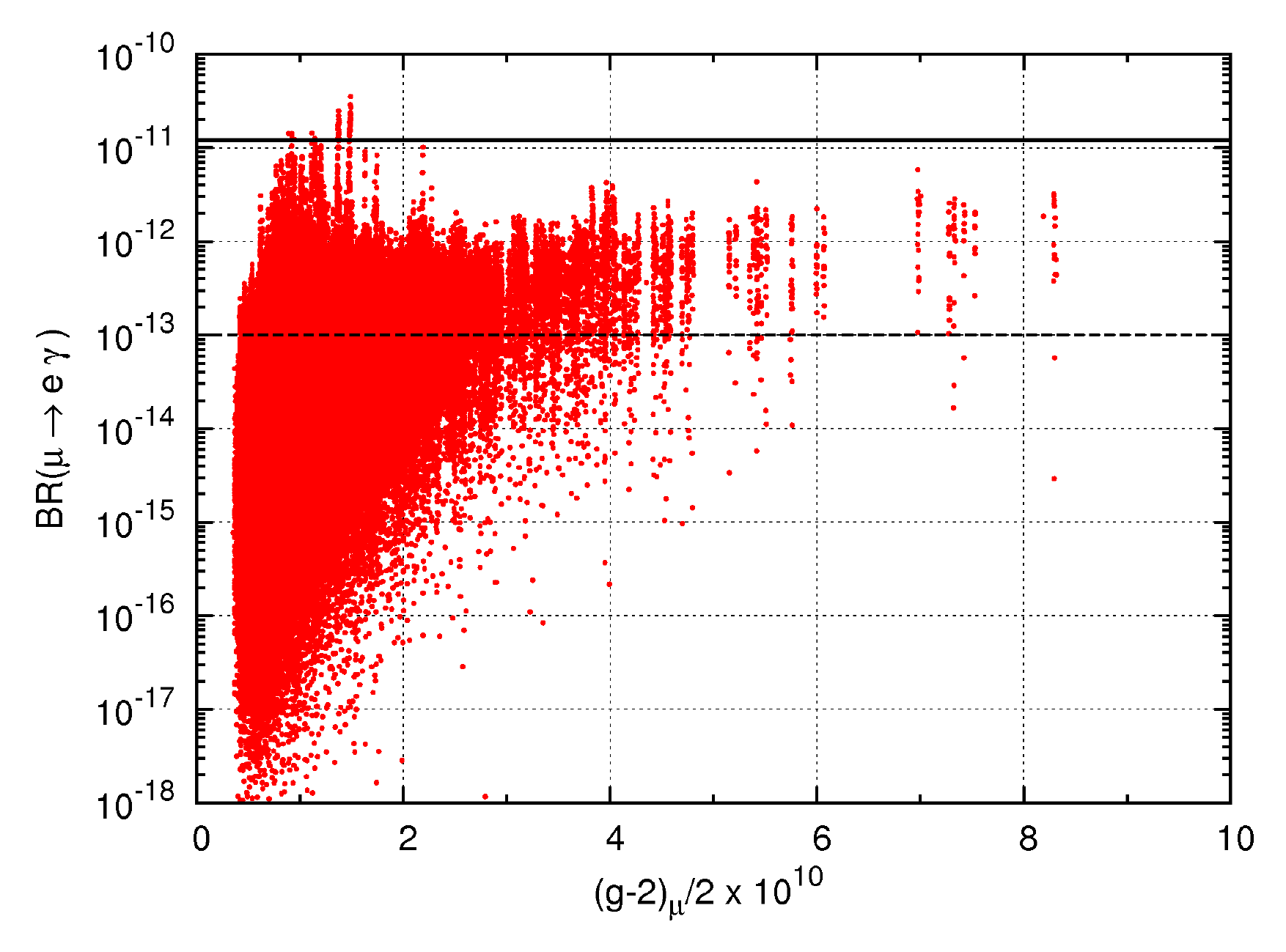}
\includegraphics[width=0.47\textwidth]{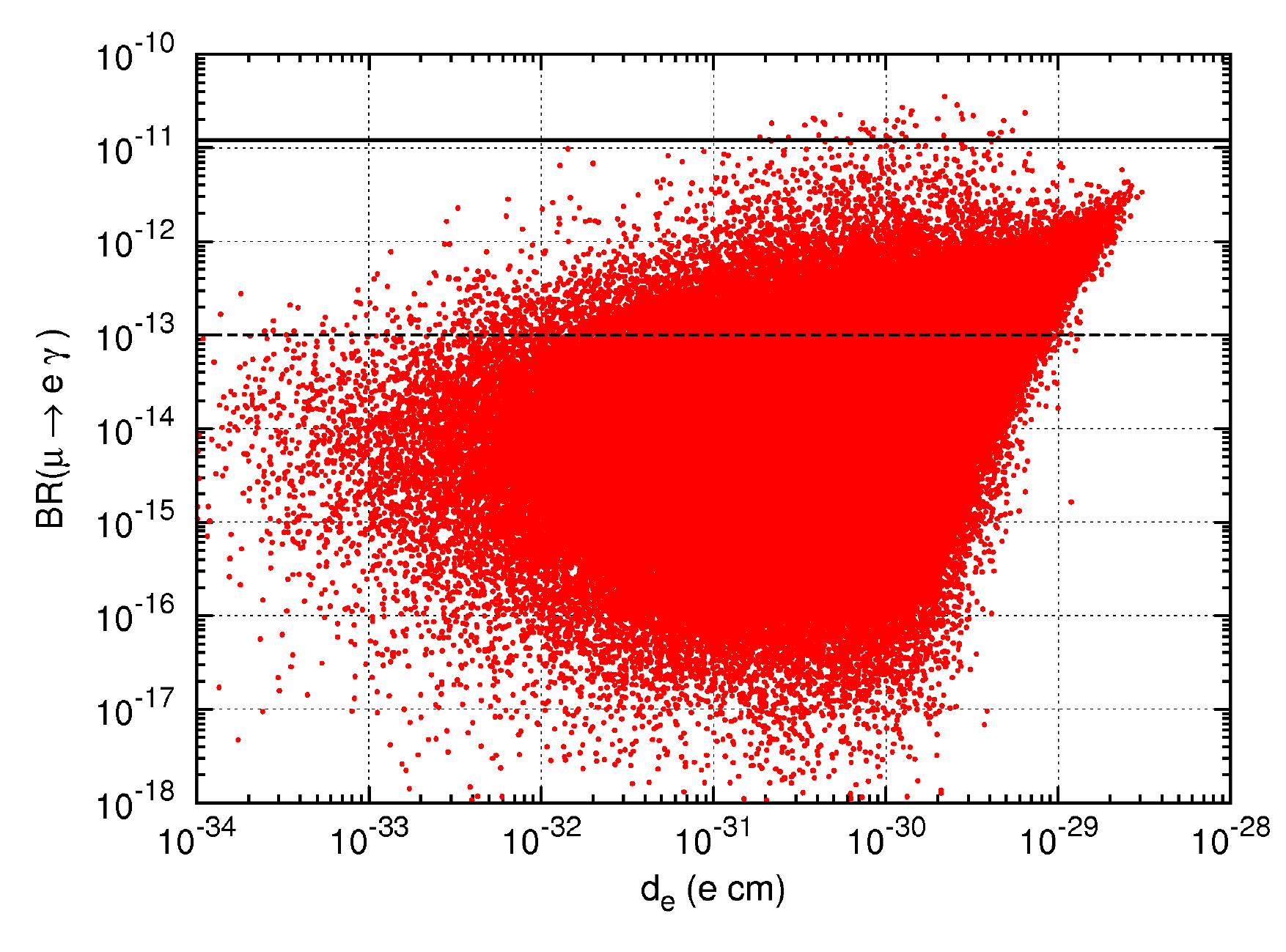}
\caption{Left: BR$(\mu\to e\gamma)$ versus the SUSY contribution to $(g-2)_\mu/2$. Right:
BR$(\mu\to e\gamma)$ versus the electron EDM, $d_e$.\label{Fig:gm2-edm}}
\end{figure}

In the left panel of Fig.~\ref{Fig:gm2-edm}, we plot BR$(\mu\to e\gamma)$ versus the SUSY contribution to the 
anomalous magnetic moment of the muon
$a^{\rm SUSY}_\mu\equiv (g-2)^{\rm SUSY}_\mu/2$. We see that no points provide $a^{\rm SUSY}_\mu \gtrsim 10^{-9}$, 
a contribution which would lower the present tension between
measurements and theoretical predictions below the 2$\sigma$ level. This is due to the heavy slepton spectrum favoured by the fit,
which clearly determines a suppression of the SUSY contribution to $(g-2)_\mu$. Within our model, a contribution at the level
required by the present tension could be achieved in set-ups with larger $\tan\beta$, e.g.\ $\sim 50$ 
(since $(g-2)^{\rm SUSY}_\mu\sim \tan^2\beta$). 

In the right panel of Fig.~\ref{Fig:gm2-edm}, we show an estimate for the electron EDM, $d_e$. 
As discussed in Sec.\ \ref{sec:estimate-fcnc}, $d_e$ receives both flavour-dependent and flavour-independent contributions. 
Such an observable then results highly complementary to BR$(\mu\to e\gamma)$ in testing the parameter space of the model,
while it does not impose at present a further constraint to parameters of the model (the current limit being 
$d_e < 1.4\times 10^{-27}$ $e$ cm \cite{CurreEDM}).
Given that the future experiments should reach $d_e \sim 10^{-30}$ $e$ cm, or below \cite{Pospelov:2005pr},
they could be able to test the model even beyond the reach of MEG, for vanishingly small values of BR$(\mu\to e\gamma)$. 
On the other hand, a portion of the parameter space in the reach of MEG should escape the next generation eEDM experiments.
However, it is interesting to notice that a large part of the parameter space should be in the reach both of 
MEG and the eEDM experiments. These two observables are therefore highly complementary and the interplay between both of them
represents a powerful test of the class of models we are discussing. In fact, an evidence of $\mu\to e\gamma$ at MEG, at the level of, say, 
BR$(\mu\to e\gamma)\sim \mathcal{O}(10^{-12})$, would predict a lower bound on $d_e$ in the reach of the future experiments.
On the other hand, observing the electron eEDM with a value of the order of $10^{-29}$ $e$ cm would imply that $\mu\to e\gamma$ 
can be observed at MEG.

\begin{figure}
\centering
\includegraphics[width=0.47\textwidth]{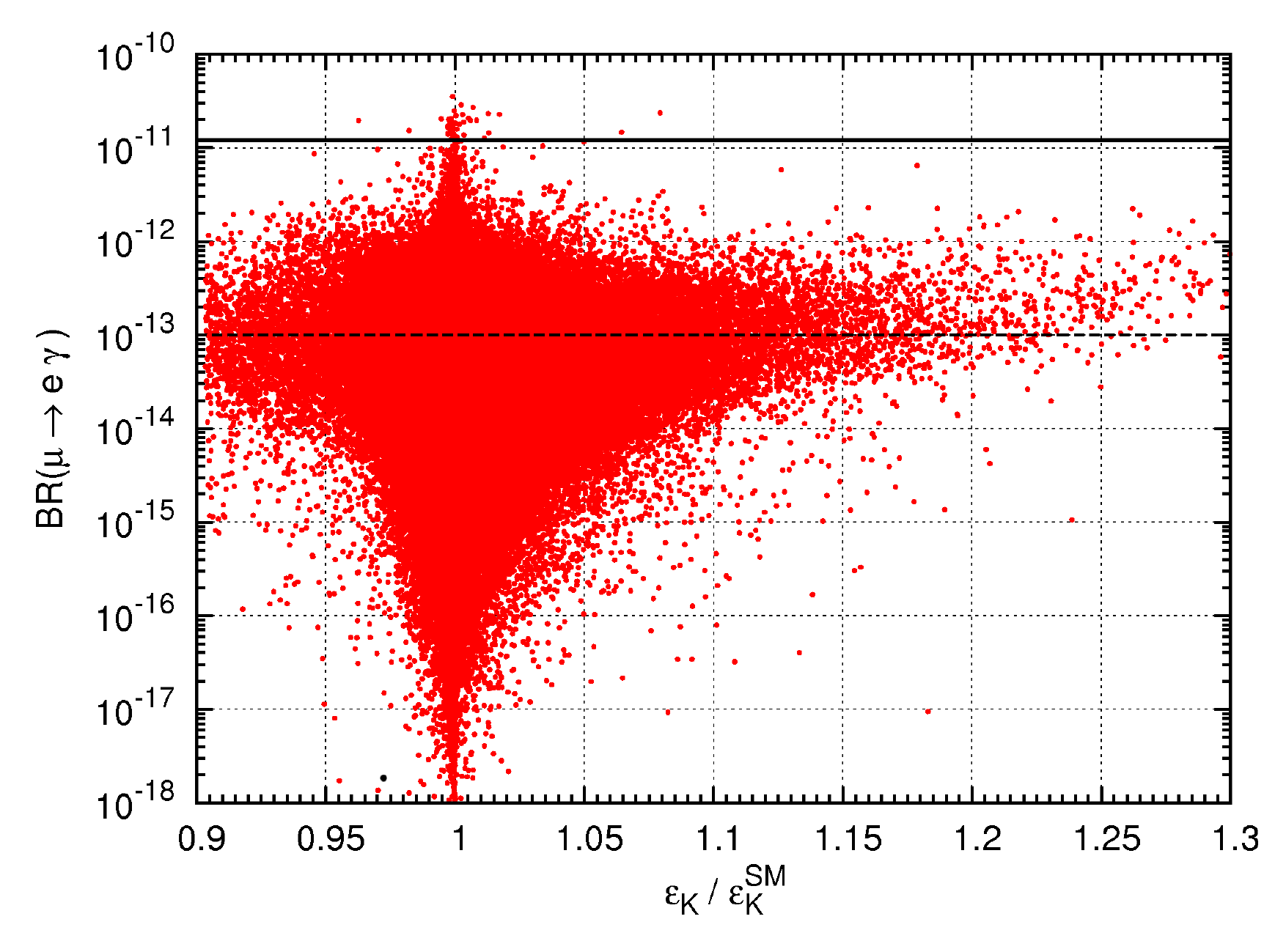}
\includegraphics[width=0.47\textwidth]{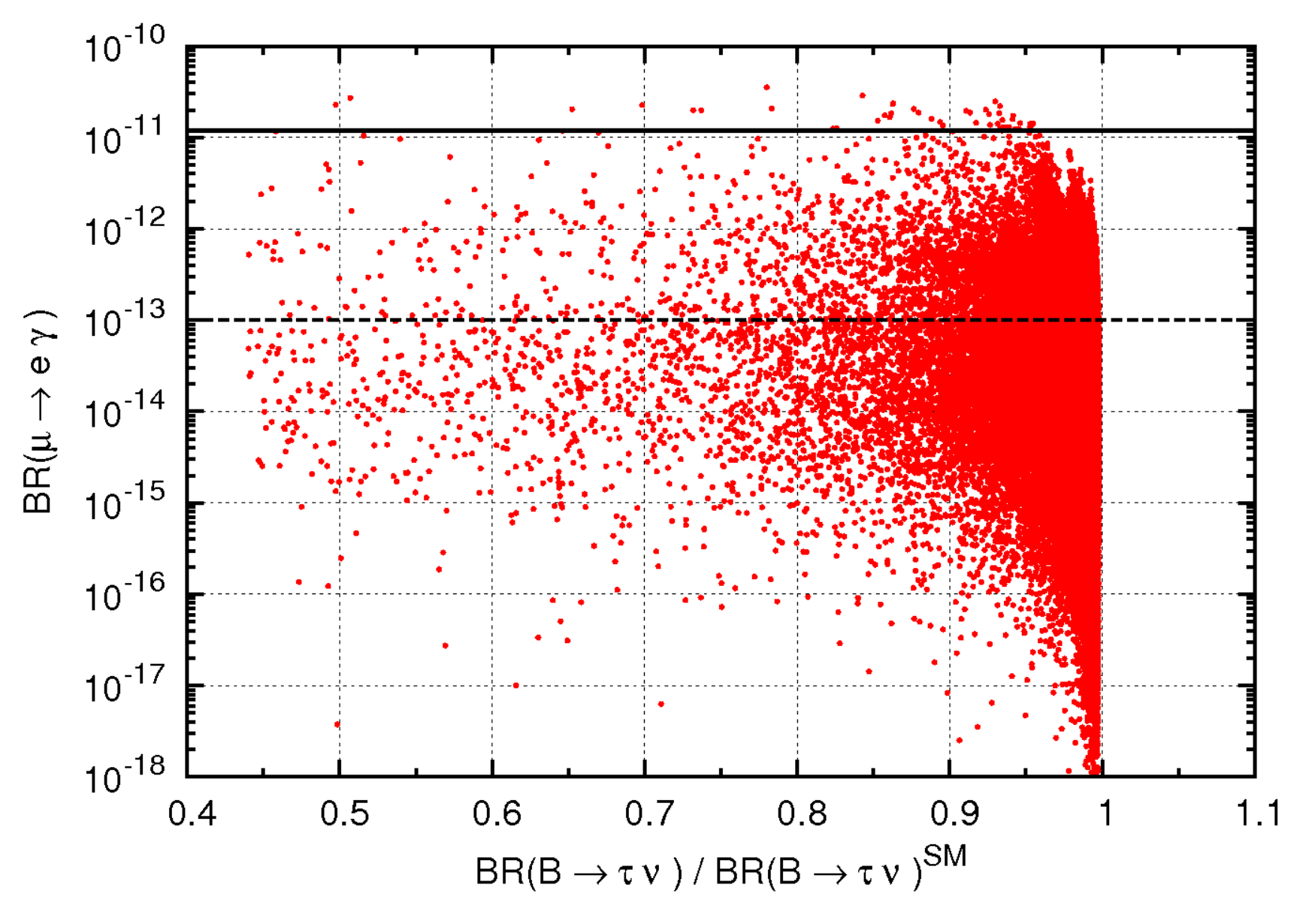}
\caption{$\mu\to e\gamma$ versus the deviation from the SM predictions of $\epsilon_K$ (left) and 
BR$(B\to \tau \nu)$ (right). \label{Fig:meg-epsK}}
\end{figure}

Recent analyses of the consistency of the Unitarity Triangle (UT) have shown some tensions among the different observables used in the fit
\cite{Soni,buras-diego}. A way to solve such tension could be a positive new physics contribution to the 
Kaon CP-violating parameter $\epsilon_K$ at the level of about +20\% \cite{buras-diego,Altmannshofer:2009ne,Buras:2010pm}. 
As discussed in the previous sections, our class of models, being embedded in a $SU(5)$ GUT, provides new sources 
of flavour violation both in the leptonic and the hadronic sector.
In the left panel of Fig.~\ref{Fig:meg-epsK}, we show the predicted BR$(\mu\to e\gamma)$ versus $\epsilon_K/\epsilon_K^{\rm SM}$. 
As we can see the SUSY contribution to $\epsilon_K$ can give sizeable deviations from the SM prediction. 
For instance, the cut of the left side of the plot reflects the fact that a large negative contribution to $\epsilon_K$ is excluded.
A contribution of order +20\% is still possible for a small portion of the parameter space.
Moreover, most of such points, which correspond to values of BR$(\mu\to e\gamma)$ below the present limit, will be tested at MEG.

$B\to \tau\nu$, such as $B_s\to\mu^+\mu^-$, might represent an important constraint for SUSY models in 
the regime of medium or large $\tan\beta$. This is due to
a potentially large charged Higgs mediated contribution, which grows with $\tan^4\beta$ and always exhibits 
the opposite sign with respect to the SM contribution,
whereas the present experimental measurement prefers values larger than the SM one (see, for instance, \cite{Altmannshofer:2009ne}).
We plot BR$(\mu\to e\gamma)$ and BR$(B\to \tau \nu)$/BR$(B\to \tau \nu)^{\rm SM}$ in the right panel of Fig.~\ref{Fig:meg-epsK}.
We see that, even if most of the points give a prediction close to the SM one,
it is still possible to have large deviations, which in future could be excluded by increasing the precision 
of the experimental determination of BR$(B\to \tau\nu)$. 
Moreover, most of the depicted points lie in the 95\% C.L. range
 $0.52 \lesssim$ BR$(B\to \tau \nu)$/BR$(B\to \tau \nu)^{\rm SM}$ $\lesssim 2.61$  \cite{Altmannshofer:2009ne}. 
This is due to the Higgs spectrum selected by our model. In fact, $m^2_{H^{\pm}}$ ($\simeq \mu^2 + m^2_{H_d} + m_W^2$) cannot be too light
since, as mentioned above, our large $A_t$ gives sizeable $\mu^2$ and $\tan\beta$ is not large enough to drive $m^2_{H_d}$ to large
negative values. Only for few points of our sample, the parameters conspire to give a quite light Higgs sector. In fact,
we find, as minimum values, $m_A\sim 261~{\rm GeV}$, $m_{H^+}\sim 272~{\rm GeV}$. These lower values correspond to points with
quite large positive values of $C_{\phi_3}$, which increase $|A_b|$ and $|A_\tau|$ and therefore contribute to drive 
$m^2_{H_d}$ negative. These points give the largest deviation from the SM
prediction for BR$(B\to \tau \nu)$ and some of them (corresponding to BR$(B\to \tau \nu)$/BR$(B\to \tau \nu)^{\rm SM} \lesssim 0.6$) 
predict ${\rm BR}(B_s\to \mu^+ \mu^-)$ at the level of $10^{-8}$, i.e. in the reach of the LHCb experiment in the near future.
However, the bulk of the points just predicts no large deviation from the SM prediction, i.e. ${\rm BR}(B_s\to \mu^+ \mu^-)\sim \mathcal{O}(10^{-9})$.

\section{Summary and Conclusions}

In this paper, we have investigated aspects of ``SUSY flavour'' models, towards predicting both flavour structures, in the context of SUGRA.
We highlighted the importance of including carefully all the SUSY-specific effects such as one-loop SUSY threshold corrections and canonical normalization effects when fitting the model to the low energy data for the fermion masses and mixing angles.
These effects entangle the flavour model with the SUSY parameters and leads to interesting predictions for the sparticle spectrum as well as, for instance, for the neutrino parameters. In addition, family symmetries introduced to explain the flavour structure of the Standard Model fermions can also make predictions testable in future flavour experiments. 

The extension of a pure model of flavour to a SUSY flavour model, and its phenomenological analysis, may be performed in the following steps:
\begin{itemize}

\item The starting point may be a family symmetry, as for instance $SU(3)$, $SO(3)$, or one of their non-Abelian discrete subgroups like $S_4$ or $A_4$ (cf.\ Sec.\ \ref{sec:Model}).

\item The discussion of the SUSY formulation of the model starts by defining the superpotential $W$.

\item The operators induced in the K\"ahler potential $K$ by a given set of (flavour) messenger fields have to be included.

\item By canonically normalizing the fields (cf.\ Sec.\ \ref{sec:normalisedYukawas}) the canonically normalized Yukawa matrices are calculated. In this way, the structure of $K$ enters the model predictions for the SM flavour parameters (i.e.\ the fermion masses and mixings).

\item From $K$ and $W$, the (canonically normalized) soft SUSY-breaking parameters can be calculated.

\item To compare the flavour model predictions to the data, the RG evolution of all parameters has to be calculated. As discussed in Sec.\ 3.5., the inclusion of the one-loop SUSY threshold corrections is mandatory for all three generations at least for moderate and large $\tan \beta$  - and connects the flavour model analysis to the SUSY parameters.

\item Finally, the SUSY flavour structure, shaped by the same symmetries which govern the SM flavour structure, can give rise to various effects testable at flavour experiments. 

\end{itemize}

We have demonstrated these issues and applied the ``SUSY extension'' to a novel class of family symmetry models,  assuming sequestering of $W$ and $K$ into visible and hidden sectors. 
The extension to SUSY, in addition to predictions for the SUSY flavour structure from the symmetries of the models, has various interesting effects: 

\begin{itemize}

\item Firstly, one-loop SUSY threshold corrections have turned out to have a crucial effect on the fit of the model to the data. Not only the ratios of quark and charged lepton masses, which are predicted by the model at the GUT scale but also the quark mixing angles, depend crucially on the threshold effects. Via them the SUSY spectrum enters into the fit of the model to the data and from this constraints on the SUSY parameters can be obtained. Using MC Monte Carlo techniques we demonstrated this for our example class of models. The main result is shown in Fig.\ \ref{fig:fullMCMCspectrum}. We also compared our full MC Monte Carlo analysis with a simple semi-analytical treatment, which may be applied as a first step to understand such kind of constraints on the SUSY spectrum from a SUSY flavour model analysis. 

\item Secondly, also the canonical normalization of the kinetic terms, which generically requires a transformation (including rescaling) of the fields in a SUGRA context, modifies the Yukawa matrices and thus affects the fit. In our class of models the canonical normalization effects were dominated by a single parameter (related to a comparatively large ratio of the flavon vev $ \phi_3 $ over the messenger mass), a typical situation in flavour models with non-Abelian family symmetries (see, e.g.\ \cite{Antusch:2007vw}). 
In our example class of models we saw, for example, that the prediction of tri-bimaximal mixing was modified by this effect, which imposed constraints on the canonical normalization parameter $\hat\zeta_3^2/2$, and predicts a correlation in the deviation from tri-bimaxmial mixing (cf.\ also \cite{Antusch:2007ib}) which could be tested in future precision neutrino oscillation experiments.

\item Thirdly, the example class of flavour models we considered and generalized here turned out to be an interesting novel class of SUSY flavour models - since the flavour structure of the soft SUSY breaking terms turned out to realise a scheme we dubbed ``Trilinear Dominance''. While the soft masses at the GUT scale have only very small off-diagonal elements, the flavour violating effects are mainly induced by the misalignment between the trilinear terms and the Yukawa matrices, governed by only a few parameters. 

\end{itemize}

We conclude that if SUSY will be discovered at the LHC, models of flavour would have to be extended to SUSY-flavour models. The testability of such models could be enhanced significantly, for example by the predictions for the  sparticle spectrum and $\tan \beta$.  Furthermore, flavour effects induced by the SUSY flavour structure could give new insights into flavour models, and characteristic SUSY flavour schemes (like ``tri-linear dominance'' discussed here) might be disentangled from others. Finally, even precision tests of neutrino mixing angles can be sensitive to SUSY/SUGRA typical  features like canonical normalization corrections, as we have seen in our example class of models.

\section*{Acknowledgments}
S.~A.\ acknowledges partial support by the DFG cluster of excellence ``Origin and Structure of the Universe''.  
M.~S.\ acknowledges partial support from the Italian government under the project number PRIN 2008XM9HLM ``Fundamental interactions in view of the Large Hadron Collider and of astro-particle physics''.

\end{document}